\documentclass[fleqn,10pt]{wlscirep}
\usepackage[utf8]{inputenc}
\usepackage[T1]{fontenc}
\usepackage[framemethod=tikz]{mdframed}
\usepackage{subfiles}
\usepackage{subcaption}
\usepackage{float}
\usepackage{cleveref}

\title{A Cognition-Affect Integrated Model of Emotion}

\author[1,*]{Sudhakar Mishra}
\author[1]{Uma Shanker Tiwary}
\affil[1]{Indian Institute of Information Technology, Allahabad, Center for Cognitive Computing, Allahabad, 211012, India}

\affil[*]{rs163@iiita.ac.in}

\begin{abstract}
The focus of the efforts for defining and modelling emotion is broadly shifting from classical definite marker theory to statistically context situated conceptual theory. However, the role of context processing and its interaction with the affect is still not comprehensively explored and modelled. With the help of neural decoding of functional networks, we have decoded cognitive functions for 12 different basic and complex emotion conditions. Using transfer learning in deep neural architecture, we arrived at the conclusion that the core affect is unable to provide varieties of emotions unless coupled with cortical cognitive functions such as autobiographical memory, dmn, self-referential, social, tom and salient event detection. Following our results, in this article, we present a 'cognition-affect integrated model of emotion' which includes many cortical and subcortical regions and their interactions. Our model suggests three testable hypotheses. First, affect and physiological sensations alone are inconsequential in defining or classifying emotions until integrated with the domain-general cognitive systems. Second, cognition and affect modulate each other throughout the generation of meaningful instance which is situated in the current context. And, finally, the structural and temporal hierarchies in the brain's organization and anatomical projections play an important role in emotion responses in terms of hierarchical activities and their durations. The model, along with the analytical and anatomical support, is presented. The article concludes with the future research questions.
\end{abstract}

\begin{document}

\flushbottom
\maketitle
%
%
\thispagestyle{empty}


\section*{Introduction}

\begin{mdframed}[frametitle={Box-1: Theories of Emotions}]
While emotion was once considered as the property of the limbic system, advances in emotion research support a more multi-faceted account which also includes activity in the neocortex in constructing emotion. Theories of emotion can be categorized in three major classes: classical/basic theory of emotion~\cite{panksepp2004affective, chapter8}, appraisal theory of emotion\cite{lazarus1980emotions, leventhal1980toward}, and conceptual act theory or constructionist theory of emotion~\cite{russell2003core, lindquist2008constructing, gendron2009reconstructing, barrett2013psychological}. 

In general, the classical theory of emotion encapsulates all the theoretical proposals associating emotions with certain bio-markers or with some evolutionarily preserved circuits mediating the emotional feelings. This universality of nature of emotion is handpicked by Paul Ekman~\cite{ekman1979facial} who reported cross-cultural similarity of facial expressions. On the other hand, the somatic-marker hypothesis by Damasio, which is originally inspired by the James-Lange theory, hypothesized the somatic biomarkers of emotions (called affect programs) whose activity was associated with the emotion eliciting cues encountered by the organism~\cite{chapter8}. Jaak Panksepp proposed seven primary affective circuits which were discovered mostly from animal models and claimed to be preserved in evolutionary process~\cite{panksepp2004affective}. In short, the classical view of emotions emphasizes characteristics such as emotions are unique mental states, emotions are caused by special mechanisms~\cite{chapter8} and specific brain circuits~\cite{panksepp2004affective}, emotions have unique manifestation on face~\cite{ekman1979facial}, voice and body state~\cite{fehr1970peripheral}, unique in responses~\cite{ekman1979facial, fehr1970peripheral}, etc.
Basic emotions were assumed to be universal in nature~\cite{ekman1971constants, ekman1969pan} and variations in emotions are considered secondary and tertiary versions of basic core emotions~\cite{panksepp2004affective}. Emotions have an evolutionary root and share the neural circuits with the non-human animals~\cite{panksepp2004affective}. This determinism in emotion processing was challenged by the researchers who proposed appraisal theories of emotions. 

Appraisal theory was proposed~\cite{arnold1960emotion} and developed~\cite{lazarus1980emotions, leventhal1980toward} to explain how different emotions may emerge from the same event, in different individuals, and on different occasions. Appraisal theories see emotions as a process (not episodes) which involves cooperation among dimensions~\cite{moors2013appraisal} including a) appraisal(evaluation of the context and subjective interaction with it to produce values for different variables), b) motivation(action tendencies), c) somatic(bodily sensations), d) motor(expressive and instrumental behaviour), and e) feeling(subjective experience). However, it is uncertain how the process of emotion operates on these dimensions. Another approach to understanding emotion as a process is that they are concepts or categories which are constructed from past experiences and beliefs just like other perceptions.

The social constructionist theory ~\cite{juslin2008all, barrett2017theory, barrett2013psychological} views emotion as a cognitively constructed social reality~\cite{barrett2012emotions} which is situated in the subjective experience of the context and subjective conceptualization~\cite{barsalou2005situated, wilson2011grounding, moffat2015situated}. This view also helped to explain the variations in the emotional experiences within an emotion category, across the subject, context and time, which is a radically different notion from the classical theory of emotion. This theory explicitly maintained the position that emotion is a socio-cultural reality. Emotions are socially and culturally adopted conceptual categories encapsulating varieties of social events and situated emotional feelings and expressions. The conceptual act theory or constructive theory of emotion encapsulates a fair amount of qualitative arguments for the role of cognitive functions in creating emotion but lacks in the quantification of contributions by these cognitive functions, the inclusion of affective processing and qualitative structural and anatomical support.

Recently proposed active inference theory~\cite{seth2013interoceptive} also supports the view that emotion is not different from perception as both are principally framed in the structural and functional asymmetry of cortical organization. The constructionist theory of emotion with hierarchical active inference model of the insular processing~\cite{craig2002opinion, bastos2012canonical}, led to the theory where emotions are interoceptive inference~\cite{seth2013interoceptive, barrett2017theory}. These inferences are drawn from the internal representations learned in the past~\cite{perri2015we, kim2017understanding, miller2018happily}. 
\end{mdframed}

\section*{Results}

We have considered online data for emotions\cite{koelstra2011deap}. The EEG and some peripheral physiological (GSR, EMG, Temperature, EOG, Respiration and Plethysmography) data were recorded while subjects were watching multimedia stimuli for 12 different emotions. The available EEG data were pre-processed with EEGLAB script\cite{delorme2004eeglab}. We have followed Makoto's pre-processing pipeline to pre-process the signal. Then the network calculation, neural decoding and MVPA analysis were done(for details see methods). 

\subsection*{\textit{Connectome and functional decoding:}}

The connectome presented in \cref{fig:connect_tom,fig:connect_saliency,fig:connect_pain,fig:connect_auto} (see also supplementary fig S5 for other functions) shows functional connections between voxel pairs which represents two connected regions. The functional connectome is based on the PLV measures on voxel-pairs. These voxels were calculated using source localization technique (for complete details, please see methods). All these voxel pairs are connected due to strong significant functional correlations (for $p<0.0002$[FDR corrected]). We have calculated set of connections between two regions. These connections have been merged by calculating the probability of significant connections (probability is calculated by dividing favourable significant connections with the total number of connections), to reduce the calculation overhead. This probability constituted the weight of connections between the pair of voxels, which represent two different regions. Further, these set of connections, across all the emotions, were utilized for functional decoding using meta-analysis based on Neurosynth database\cite{yarkoni2011large}. Error, cognitive control, and conflict-related voxels coordinates were combined to form saliency network(fig~\ref{fig:connect_saliency}). It is based on the literature\cite{ham2013cognitive, menon2010saliency, purcell2012supplementary, wyble2008strategic} and overlapping connections shown in table-S4 (supplementary section) which shows that the cognitive control, error, and saliency networks are overlapping relatively more among each other than with the other functions.

Neurosynth database was utilized for decoding terms associated with each voxel pair. Neurosynth provides the meta-analysis of many brain functions by analyzing the vast amount of research articles (approx. 14371 studies)\cite{yarkoni2011large}. For each calculated functional connections between voxel pair, the co-activated set of voxels within the 5mm distance cut-off was included. This step was done to compensate the source localization error using sLoreta, which is reported to be approximately 1.45mm($\pm$3.71mm)\cite{song2015eeg}. The resultant co-activation map was decoded for the associated terms along with the posterior probability of these associated terms(or functions, used here interchangeably) based on the association test (FDR corrected with the expected 0.01). For all the connected voxel pairs, the associated term with posterior probability more than 0.3 were considered. 18 terms with some subcategories and overlaps among each other (see Supplementary table-S4 for overlap and \cref{fig:connect_auto,fig:connect_pain,fig:connect_saliency,fig:connect_tom,fig:statmap_auto,fig:statmap_pain,fig:statmap_saliency,fig:statmap_pain} for functional connections among and activation of different brain regions) were considered for the MVPA classification using transfer learning and convolution neural network approach(fig~\ref{fig:Deep_Model}). For these functions, the meta-analytic co-activation map were extracted from neurosynth database and plotted along with the axial or transverse plane(see statistical map in \cref{fig:statmap_tom,fig:statmap_saliency,fig:statmap_pain,fig:statmap_auto}). This way of calculation, to some extent, reduced the limitation of source localization up to cortical layers and allowed to consider structures from subcortical regions too, which were essential to propose the model. The included subcortical structures in the model were based on meta-analysis of valence, arousal and pain(physiological sensations). These three terms were considered as representing affect and affective sensations. The subcortical structures which are active for valence, arousal and physiological sensation are the amygdala, putamen, cerebellum, caudate nucleus, thalamus, pallidum, vermis and hippocampus (Please see supplementary section 'Active regions for valence, arousal and pain'). 

As shown in the connectome(\cref{fig:connect_auto,fig:connect_pain,fig:connect_saliency,fig:connect_tom,fig:statmap_auto,fig:statmap_pain,fig:statmap_saliency,fig:statmap_tom}), we are getting some regions which are mediating the peripheral connections and working as hubs. These regions include precuneus(7), PCC(31), ACC(32),  MidTG(21), IPL (40), PreG(6), STG(22) and LG (18). These hubs are reported in the domain-general systems which mediate multimodal information among brain regions\cite{power2013evidence}, between uni-modal processing regions and sub-cortical structures\cite{tomasi2011association} and thus contribute in cognitive construction\cite{barrett2013psychological, power2013evidence}.
Mostly, we are getting functional connectivity of precuneus with all other parts of the brain. The role of precuneus is implicated in autobiographical memory-related tasks\cite{barry2018neurobiology}, in the tasks related to first-person perspective\cite{cabanis2013precuneus}, mental imagery and navigation\cite{thornton2019social}, conscious agency\cite{cavanna2007precuneus} and also in the resting state condition\cite{li2019developmental}. Precuneus is reported to have widespread connections from several cortical and subcortical regions which includes lateral parietal regions, lateral and medial frontal cortex, temporal pole, temporo-parieto-occipital, thalamus, and brain stem nuclei\cite{cavanna2006precuneus}. To our knowledge, no study has reported the connection of precuneus with sensory cortices directly, but it has strong functional and structural connections with associative regions, which makes it the site for higher-order processing integration.

\subsection*{\textit{Deep Neural Network based MVPA analysis using transfer learning:}}

\begin{figure*}[ht!]
\begin{subfigure}[t]{0.38\textwidth}
\graphicspath{{pictures/}}
\includegraphics[width=\textwidth, height=1.25\textwidth]{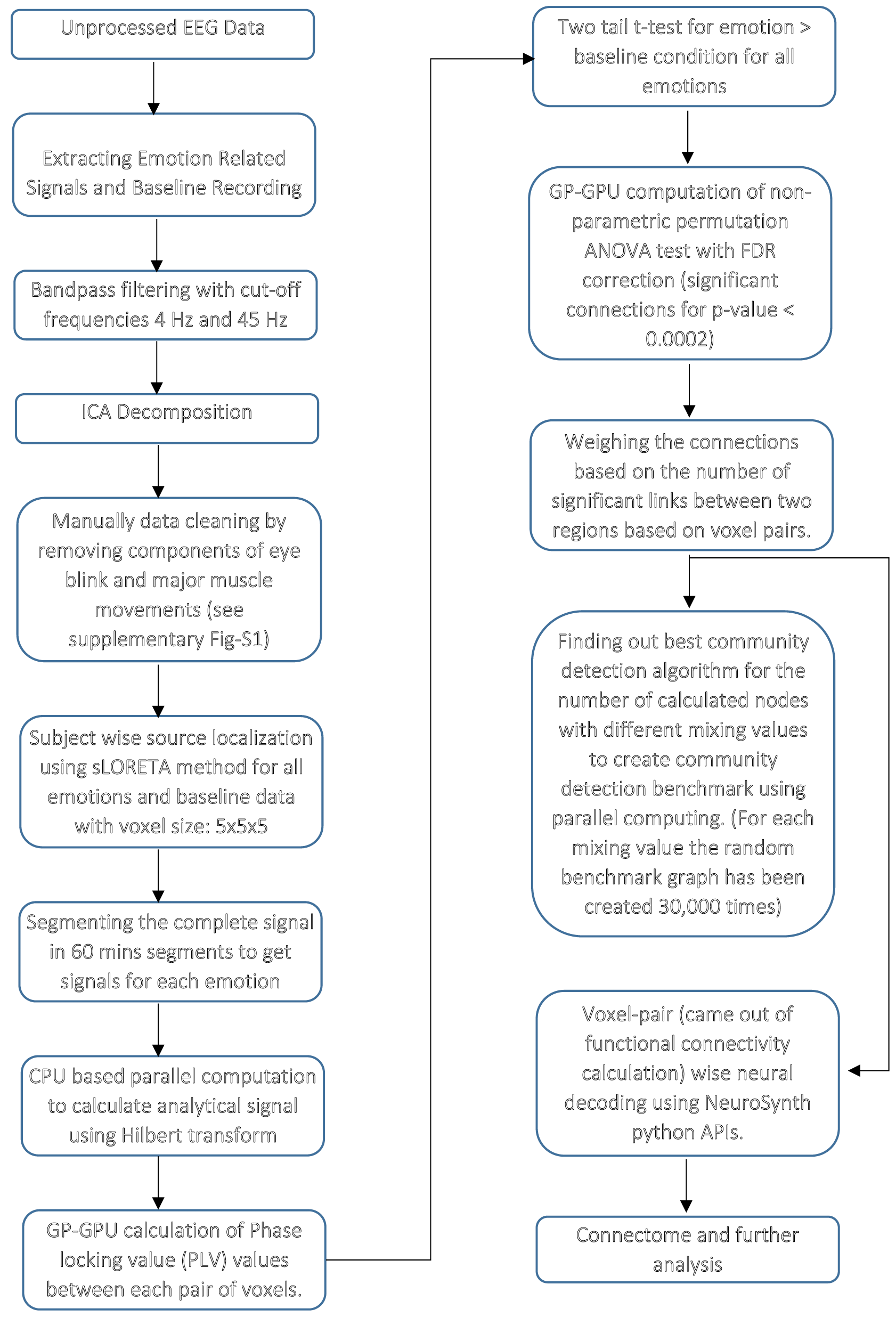}
\caption{} 
\label{fig:Methodology}
\end{subfigure}
\hfill
\begin{subfigure}[t]{0.38\textwidth}
\centering
\graphicspath{{pictures/}}
\includegraphics[width=0.6\textwidth, height=1.25\textwidth]{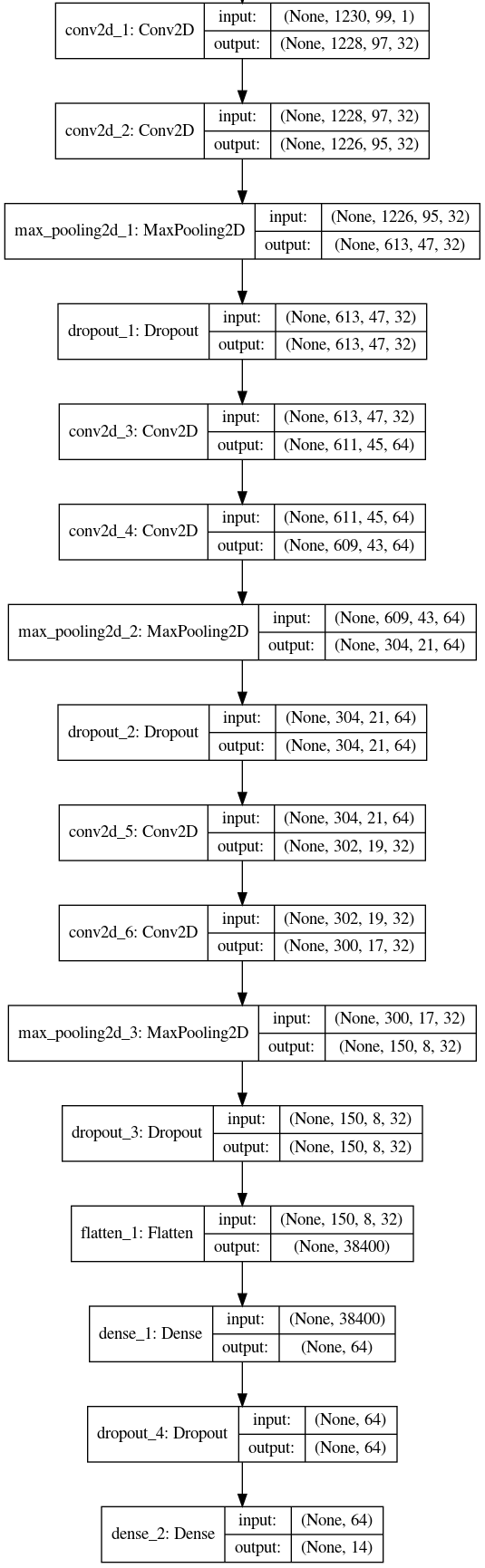}
\caption{} 
\label{fig:Deep_Model}
\end{subfigure}
\hfill
\begin{subfigure}[t]{0.2\textwidth}
\graphicspath{{pdf/}}
\includegraphics[width=\textwidth, height=1.25\textwidth]{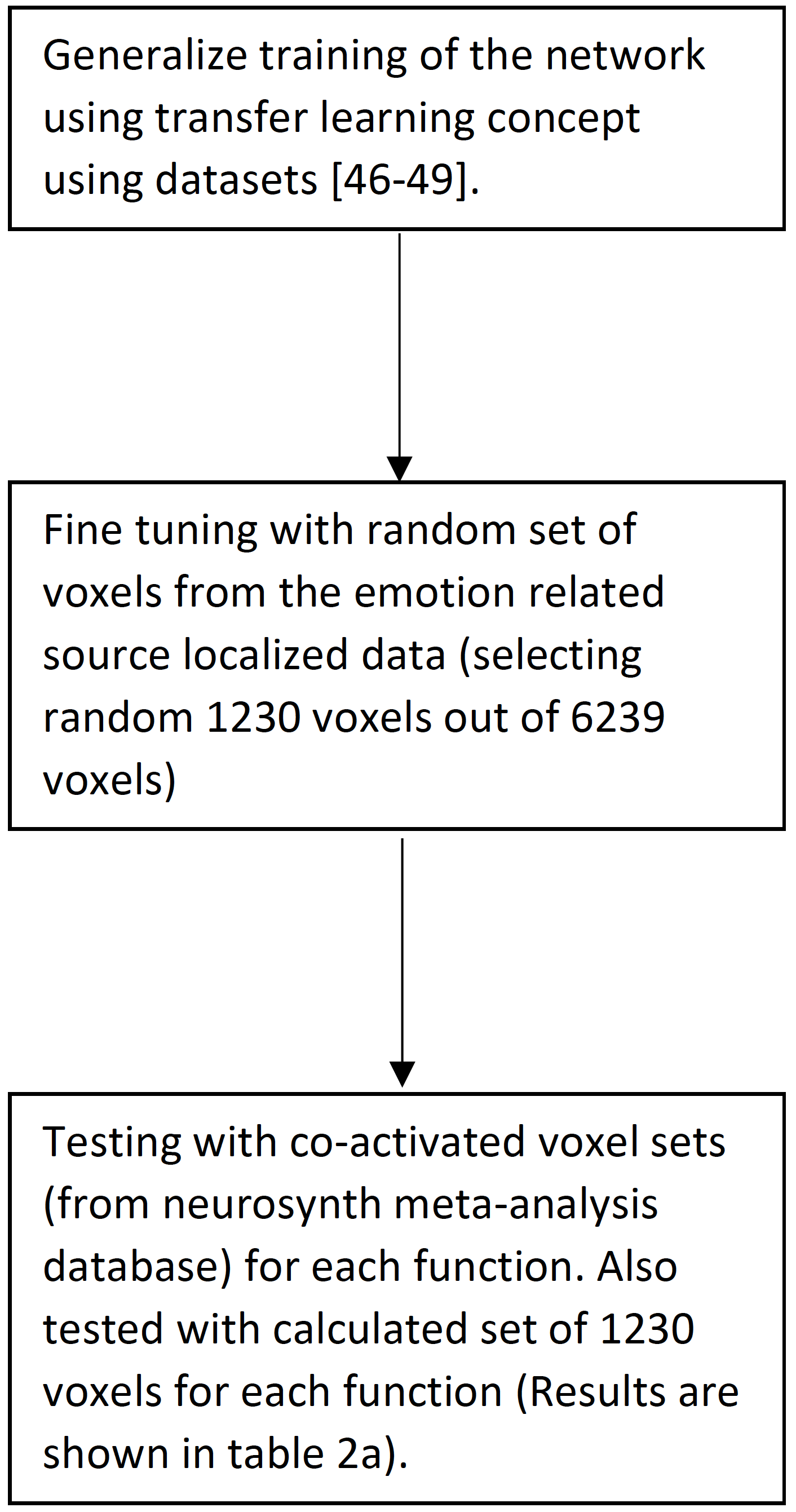}
\caption{} 
\label{fig:TL}
\end{subfigure}

\caption{\textbf{Methodology Flow Chart and Deep Learning Architecture: }(a)Methodology is reflecting the flow of complete analysis. (b) The summary of deep learning architecture having six convolution layers, three max-pooling layers and four dropout layers. The categorization is done using dense connected neural network layers after getting the self-learned feature representations at the flatten layer. (c) Flowchart for transfer learning. Preconditioning weights of deep learning architecture with different EEG datasets and fine-tuning the architecture for final testing. }
\label{fig:methods}
\end{figure*}

The 18 decoded functions (table in fig~\ref{fig:mvpatable}) were considered for the MVPA classification using transfer learning and convolution neural network approach(~\cref{fig:Deep_Model,fig:TL}). We utilized transfer learning\cite{yang2013theory, pan2009survey} for training the deep learning model since we had limited data. To train the model(~\cref{fig:Deep_Model,fig:TL}), we utilized four different kinds of data from the physionet databank. First, EEG data on mental arithmetic task\cite{zyma2019electroencephalograms}, second, EEG data for motor movement and imagery\cite{schalk2004bci2000}, third, ERP based BCI recording on target and non-target set of characters\cite{citi2010documenting}, and fourth, MAMEM Steady-State Visually Evoked Potential(SSVEP) EEG database\cite{oikonomou2016comparative}. All these datasets had different categories to be classified. Our main intent behind taking all these different EEG datasets is to let the model get familiarized and set its parameters for the general characteristics of electrophysiological signals by learning filter kernels and transfer this learning to a new task.
The concept of transfer learning resembles the human learning in a way that humans see many examples on moment to moment basis(for example, images) and get trained for separating two objects based on their general features only(for example, edges, corners, blobs, textures and so on) even without knowing specific name of these objects\cite{brown1988preschool}. Before training the model, EEG channels were source localized to find out source voxel activity using sLORETA. Activity in 6239 voxels was calculated using source localization method. One of the limitations of transfer learning is that the input layer should have the same number of nodes. For the final testing, we had only 1230 voxels(these 1230 voxels are as per our functional connection analysis), and that's why we made chunks of 1230 voxels during training so that we have the same number of nodes in the input layer. The selection of these voxels, during training, was done randomly. In this way, we created 200 batches of training input for each of the considered datasets (the batch mentioned here should not be confused with deep learning batch size which is normally used during training to deal with the computational and convergence time). That means, a total of 800 batches of training input with varying number of samples. Machine trained on one batch was used in the next batch and so on. In this way, the general-purpose EEG model, which we trained, learned very general features and structures of EEG. The trained model was then fine-tuned with the set of 1230 voxels randomly picked from our source localized data with 6239 voxels (although for these random set of voxels we had very less accuracy, we succeeded in tuning the model for the brain signals for emotion stimulation.). The fine-tuned model was applied in a testing set of data with the limited number of samples from the emotion experiment. 

Based on our results, we claim that the higher-order functions like salience detection, autobiographical memory(including self-referential), theory of mind/social and default mode network are crucial for providing meaning and thus classifying the affect in the frame of context and making the emotion a reality. Affect or physiological sensation could not classify 12 different emotions (namely happy, fun, exciting, love, lovely, mellow, hate, melancholy, sad, sentimental, shock , and terrible) with more than a chance accuracy of 31\%. Although, they might be contributing in salient feature selection through long-range projections (see fig~\ref{fig:connect_pain} \& \ref{fig:Layer_1_Projections}). The presented computational analysis provides strong support for the notion that emotional experiences are significantly contributed by meaning-making domain-general cognitive functions. The association of affect with the context can be either conscious or at the subliminal level. This
association could be implicit with the contribution of the physiological neural map during very early sensory processing (for example, projections of the physiological neural map onto sensory cortices\cite{abs2018learning}). And, it could also be delayed with the delayed interpretation of the complex social situation (for example, model of the mind, in which case the context interpretation will be more explicit\cite{preckel2018interaction}).

\section*{Discussion}

Earlier model of emotion were more inclined towards classical and deterministic aspects of it~\cite{panksepp2004affective,ekman1979facial,fehr1970peripheral}. With the continuous development in understanding~\cite{lazarus1980emotions, leventhal1980toward,russell2003core, lindquist2008constructing, gendron2009reconstructing, barrett2013psychological}, later proposed models started modeling emotion as a non-deterministic and distributed phenomena. Following the same line of development, we have proposed here a conceptual model of emotion supported by the analytical observations of long-range cortico-cortical and cortico-subcortical coactivations, calculated cognitive functions, and description in the framework of brain's organization (functional asymmetry, the laminar organization, microscopic descriptions). As the laminar organization \cite{von2009cellular, barbas1997cortical} and functional asymmetry \cite{d2016recruitment} are reported to be the general structural and functional organization of the brain, we considered this concept for the processing of emotions too. 

The presented layered model (see fig~\ref{fig:model_}) signifies the interaction between affect and cognition in loop to construct an event of emotion. Based on our results and the model, we suggest that communications among different brain regions, which are responsible for social context and self related event processing (for example places, objects, goals and so on), salient feature detection, attention, reward/punishment, hedonic value, and physiological sensations (all discussed separately in the next section), takes place to create an event of emotion. 

Our model explicitly explains the nature of emotion against the universality of it in a way that emotion itself is part of the process underlying on the brain's dynamic connectivity organization. And, these dynamic interactions construct an affective subjective experience which is called an emotion. Our model also argues beyond the concept of appraisal model in a way that emotion is not merely reaction to the appraised stimulus but encoded in experience. Our model is inferred from calculated cognitive functions (using neural decoding and MVPA analysis) rather than speculative arguments on the involvement of different cognitive functions unlike in social constructionist model of emotions.

Using MVPA analysis, we found that affective sensation plays a little role in categorizing emotions and they have to be contextualized with the domain-general and socio-cultural processing(table in fig~\ref{fig:mvpatable} \& fig~\ref{fig:Connectome_Plots}) to categorize emotions. Emotions have no distinct and defined clear boundaries based on physiology, expressions or anatomy but fuzzy and statistical in nature and they are learned. Rather than characterized with specific biomarkers, emotions can be categorized based on MVPA analysis with statistical representations which varies within(less) and among(more) different categories of emotions. These statistical representations are learned contexts for an emotion. These learned representations encode various levels of detailed and abstract interoceptive and exteroceptive information~\cite{colgin2009frequency} of subjective importance across the subcortex and cortex with behavioural goals during learning (learning due to sufficient supervised or unsupervised encounters) all the way up to highly generalized, amodal\cite{ravassard2013multisensory} and abstract concepts\cite{milivojevic2013mnemonic}. When any cue (internally predicted and/or externally presented), which is informative enough to activate the onset of an episodic event, is encountered, the sequence of events in time\cite{macdonald2011hippocampal, johnson2007neural} within the delineated spatial boundary\cite{solstad2008representation} of environmental context (altogether creating an integrated state of emotional episode~\cite{milivojevic2013mnemonic}) is recalled sequentially in order. These contextual cues are encoded in the chunk of neural assemblies\cite{pastalkova2008internally} and maintained in the cortical hierarchy\cite{milivojevic2013mnemonic}. These sequence of events might be fulfilling different subjective behavioural goals which are the result of the con-specific social arrangement for better human survival in a culturally created environment. This dynamic, culturally created environment causes the development of cognition throughout life (due to the life long interaction of an individual with the uncertainty in the environment)\cite{pinker2010cognitive}. This ongoing learning of socio-cognitive perception shapes affects into the spectrum of emotions. With cognitive learning, affects find their socio-culturally defined meaning, representations and expressions. In this way, the core affect is nurtured over its innate nature to maintain the allostatic stability of the organism. The culturally agreed and nurtured affective representation is remembered for foraging the contextually situated sequence of events in the future when a similar context is encountered. 

Follow-up text describes different submodules contributing significantly in constructing an emotion. It includes salience network, social/ToM, autobiographical memory/self-referential/dmn, affect and interoception. We also discuss the integration of information within and between subcortical and cortical systems along with the consideration of the structural layout and anatomical projections. Due to lack of space some other details regarding the below discussed functions, interactions and anatomical projections are included in the supplementary section 'Decoded cognitive functions from the functional connections'.

\subsection*{\textit{Social Processing and Theory of Mind:}}
Evaluation of social context is distributed among regions which are responsible for the spatial, object, face, and temporal sequence processing. These constituting elements of the environment work as a cue to retrieve the concept which can be positive or negative based on the associated reward and social gain. Reward, attention, and physiological affective sensations modulate the recurrent circuits (fig~\ref{fig:Granular_Agranular}) to facilitate salient feature detection at the different levels of representation hierarchy.(fig~\ref{fig:HierarchyAnatomy}). Social and emotional behaviour are intertwined, which can be broken in the series of processing constructs\cite{beer2006social}. Recognition of intention(ToM) and acquisition of social-emotional value leads to the modulation of low-level affective and physiological activity. The low-level activity, in turn, causes changes in higher-order cognitive processing and representations, which leads to conceptual inference about emotion. For example, decoding other's intentions as harmful may cause intense physiological activity and release of stress hormones. This affective modulation influences the cortical processing\cite{schmitgen2016stimulus} and representations which amount to infer the concept of fear emotion in the current situation. So, the emotion itself is a subjective meaning projected in the social context that is contributed by the subjective, physiological saliency. The context is being provided by large-scale brain networks. These contextual representations are acquired in the service of allostasis\cite{sterling2012allostasis} regulation to meet the demand of the situation in the socio-cultural environment. For example, a fight or flight response in the above situation. 

Emotions are also learned as a concept and social phenomena which are to extract the optimal benefit of interest to self and/or others. The emotion learning gets matured with the development of  long-range connections and dendritic arborization\cite{srivastava2012social} (fig~\ref{fig:Microstructure}). An increasing number of dendritic spines supporting long-range and short-range feedback projections\cite{zagha2013motor, averbeck2017unbelievable, petro2014contributions, wu2005cooperative, happel2014dopamine, gonzalez2003dopamine, abs2018learning, seeman2018sparse, petreanu2009subcellular, zhang2014long, jinno2007neuronal, wang2019anterior} witness these optimal representations. These feedback (top-down) projections (fig~\ref{fig:Layer_wise_communication} \& \ref{fig:Microstructure})modulate the intermediate processing\cite{d2016recruitment}, sensory processing\cite{zagha2013motor, zhang2014long} and actions\cite{zagha2013motor} as per the internal representations of the exact or similar situations\cite{quiroga2012concept} which are experienced in the past. Previous studies has reported activity in temporal pole(38), IPL40, IFG47, CinG32 in affective ToM; insula in integration of cognition and affect(stimulus valence)\cite{gu2012cognition, mier2010involvement}. Activity in these regions is found in our results(fig~\ref{fig:connect_tom})[The regions with the significant activity based on neurosynth meta-analysis database is included in supplementary section table S6 \& S7]. Other than these regions, we have also observed activity in the precuneus, which is not explored and attended seriously. We suggest that precuneus plays an important role in cognition-affect interaction. Since precuneus is also designated as a hub and communicate with other associative cortical regions, it might be orchestrating the cognition-affect interaction.

\subsection*{\textit{Autobiographical or episodic memory and concept cells:}}
The episodic event and concept are acquired with its physical and temporal structure in the environment. Activation of an internal representation of specific context can activate neural patterns down the path and modulate regions related to core affect eg.basolateral amygdala (BLA) and central amygdala (CeA)\cite{ramirez2013creating}. Using anatomical labelling of regions, which are calculated as consistently active (in neurosynth meta-analysis) for autobiographical memory(AM) and core-affected regions, we observed that some regions including bi-GRe, bi-Hipp, bi-PIns, bi-PHG, rTMP and bi-MTG are overlapping for these two functionalities(abbr. are in supplementary table S7). We also observed overlapping in cortical functional connections in our functional connectivity analysis which includes Pre7, Pre31, ACC32, IPL40, MTG21 and CinG31. We considered dmn, AM and self-referential as one group here since in our results we are getting maximum overlapping (see table-S4 in the supplementary section) as well as they are reported to be closely related in the literature\cite{kim2012dual}. 

As shown in the model(see fig~\ref{fig:model_}), MTL system is comprised of several regions including the hippocampus, entorhinal cortex, perirhinal cortex and parahippocampal cortex. The parahippocampal and perirhinal cortices receive direct inputs from cortical sensory areas and send this information to the entorhinal cortex, which, in turn, projects to the hippocampus(top of the MTL hierarchical structure). The hierarchy of processing in different regions of MTL is presented in terms of selectivity, response latency, response units, and modality-specific invariance and encoding. Indeed, there is an increase in selectivity of neurons across the MTL, with the lowest selectivity found in the parahippocampal cortex and the highest in the hippocampus\cite{mormann2008latency, ison2011selectivity}. Communication between the MTL system and sensory regions activate older memories for the perception of new events. The abstract memory representations, in terms of concept, in the MTL system and local hierarchically ordered complex feature representations in the cortical regions, interact to construct the whole brain event.

Space, time and contextual details are represented in hierarchical manner\cite{milivojevic2013mnemonic} in terms of different degrees (general and abstract to specific and implicit details) of spatial\cite{pastalkova2008internally, buzsaki2013memory}, temporal\cite{pastalkova2008internally, buzsaki2013memory} and contextual details\cite{buzsaki2013memory}, respectively. The higher level allocentric representation of space as an internal spatial-map, information related to temporal gap between the sequence of events\cite{johnson2007neural, macdonald2011hippocampal}, and integrated concept representations of mnemonic items (such as people, objects, and landmarks)\cite{wikenheiser2015hippocampal} are mostly reported to be encoded in place and grid cells, time(encoded in the order of activation) cells and concept cells, respectively, in the MTL system\cite{eichenbaum2014can}. It is evident with the place cells\cite{solstad2008representation, pfeiffer2013hippocampal}, grid cells\cite{solstad2008representation} and concept cells that abstract or coarse level information is represented in exclusive manner and combines many low-level representations or events\cite{eichenbaum2014can} (for example, low-level affective and sensory details). To represent medium-level(entities in the scenes) and low-level(entities specific details), the memory network extends to interconnected cortical regions such as EC, perirhinal, mPFC, PPC, mPC, lateral temporal cortex, insula and so on (as shown in \cref{fig:connect_auto,fig:statmap_auto,fig:regionheapmap,fig:model_}). This memory network encodes domain or function-specific memories which can range over the specific to abstract scale(for example, unimodal to amodal). Other than the structure, an alternative way of understanding this hierarchy is what is the duration of change or frequency of change in the neural activity in a particular region(fig~\ref{fig:HierarchyAnatomy}). For example, in the case of auditory modality, due to change in phonemes and letter the neural dynamics in the auditory cortex get changed. Whereas, cortical columns in pSTG or TPJ may respond to a word in which temporal frequency of change is less than the frequency of change in the auditory cortex and so on\cite{hasson2015hierarchical}. The temporal hierarchy is also responsible for variation in emotion response considering the implicit to abstract level processing in the hierarchy. The encoded experience in the spatial and temporal hierarchy is reinstated to construct a predicted event(e.g. emotional event) which is anticipating the representation and sensations in the sensory organs\cite{jai2018specific}.

\subsection*{\textit{Salience processing and attention:}}
Salience processing can be influenced by minimally conscious physiological states, goals and drives(bottom-up), on the one hand, and conscious effect of previous experiences, goals and memories(top-down), on the other. In the former condition, visceral and autonomic activity, and physiological homeostasis condition can influence what is perceived to be salient.  In the latter condition, conscious salience processing can be goal-directed and dependent on top-down attention and cognitive control processes. In the goal-directed salience processing, salience network also includes ACC and PCC/Precuneus(see fig~\ref{fig:connect_saliency}), which are the reported sites for attention\cite{benau2018blink}.  The functional coactivation pattern between salience network, pain, valence and arousal includes overlapping region such as bi-ACgG, bi-AIns, bi-MFC, bi-MTG, bi-PIns, rPut, and lPHG as shown in the fig~\ref{fig:statmap_saliency}(abbr. are in supplementary table S7).

The salience network (which also includes interoceptive information) plays a vital role in emotion, cognition, and perception. The thalamic nucleus is a sensory relay station which receives lamina-1 axons and projects majorly to the primary interoceptive cortex and minorly to the somatosensory cortex. The interoceptive signals are further received by decreasing granular sites (mid-insula and anterior insula in less differentiated structural order, respectively) where integration with other input modalities takes place\cite{ad2015you}. The salience network involves anterior insula since at anterior insula different input modalities integrate\cite{farb2015interoception, deen2010three} and projects output signal about the subjective physiological significance to other parts of the cortex\cite{duquette2017increasing}. The anterior insula also has the mirror neurons, which can mimic the structure and motion of the physical world and embodies it in the physiological processing. Salient event is distinguished from non-salient events in a way that it has significance for subjective well-being\cite{benau2018blink, farb2015interoception, duquette2017increasing, thoits2013volunteer}. The salient information about subjective well-being and allostasis stability plays a decisive role in attention, cognitive control and error processing (table in fig~\ref{fig:mvpatable} \& supplementary table-S4). Subsequently, these systems via long-range connections, modulate the functional microcircuits in other brain regions and contribute in feature selectivity(fig~\ref{fig:Layer_1_Projections})\cite{wagatsuma2011layer, zhang2014long} as per the top-down and bottom-up saliency. 



\subsection*{\textit{Affect, Physiological State of the Body and Affect-Interoception-Context Interaction:}} 
There is plenty of evidence that the subcortical input (thalamoamygdala\cite{luo2007neural},  thalamocortical\cite{schwartze2012temporal}) provides preliminary input to create coarse information based mental representations encoding the expectation of a series of actions. Subcortical nuclei, like amygdala and pulvinar, directly interact with the dorsal stream and frontoparietal attention network. Amygdala also modulates the response of ventral visual stream, orbitofrontal and anterior cingulate cortex. Other than the direct connections, amygdala, accumbens, pulvinar, and superior coliculus modulate the cortex by modulating brain stem nuclei\cite{bertini2018pulvinar, tamietto2010neural, williams2005differential}. Tracing of the temporal structure of acoustic events reveals the role of thalamocortical connections in the automatic encoding of event-based temporal structure with high temporal precision, whereas the striato-thalamocortical connections engage in the attention-dependent evaluation of longer-range intervals\cite{schwartze2012temporal}. Sub-cortical and lower-level perception of emotional stimuli is reported when given the audio-visual stimulation\cite{tamietto2010collicular, tamietto2012subcortical} and these subcortical coding of relevant changes in audio-visual signals as event markers may be facilitating the activity in the cortical hierarchy. For example, for the visual modality in blindsight patients, activity in the amygdala, pulvinar, and superior colliculi takes place\cite{tamietto2010collicular}.
The early-stage activity in amygdala and hippocampus causes modulation of later stage cortical activities (for example, feedback modulation to sensory processing). 


Different social context \cite{wiech2016deconstructing}, with the same amygdala neural ensemble \cite{corder2019amygdalar}, can cause different emotions. On the other hand, the same context with different amygdala activation might evoke the different intensity of the same categorical emotion as the different intensity of stimulus can evoke different neuronal ensemble in the amygdala \cite{corder2019amygdalar}.
For example, basolateral amygdala(BLA) mediates associative learning for both fear and reward\cite{kim2017encoding, maren2017synapse, ressler2019synaptic, lucas2016multimodal}. Different BLA projections to the nucleus accumbens(NAc), medial aspect of the central amygdala and ventral hippocampus \cite{beyeler2018organization} distinctly alter motivated behaviour. In the functional microcircuits(fig~\ref{fig:Granular_Agranular}), some neurons get excited and some get inhibited composing neuronal ensemble for valence (positive or negative) conditioned stimulus \cite{beyeler2018organization}. Within the BLA, the neural responses to cues that predict rewarding(BLA-NAc projecting neural population and BLA-vHPC) and aversive(BLA-CeA projecting neural population and BLA-vHPC) outcomes differ depending on the anatomical projection target of each neuronal sub-population \cite{beyeler2016divergent}. 

A given neural module may not be permanently dedicated to just one affective function, but it may have multiple affective modes as states or conditions change\cite{berridge2019affective}. These affective states/conditions facilitate affective modules with different affective valence-related functions. The affective modules(for example, approach, avoidance, reward, and punishment) are the functions of the neuro-biological patterns which are represented in the brain as the global context, internal physiological state of the body and the pattern of the electrical excitation in terms of frequency and time within the module. These patterns influence the affective mode of the module. Affective valence is a generated response which depends on the situations and the interplay between subcortical and cortical processing structures\cite{berridge2019affective}. Most sites within the nucleus accumbens medial shell and amygdala are not permanently tuned to one affective valence function but rather have multiple modes that can dynamically flip to generate motivation for opposite affect as the conditions change\cite{fadok2018new, berridge2019affective}. Central amygdala (CeA) output signal can give rise to very different behavioural responses depending on the functional states of other brain areas reflecting external and internal factors, such as context, anxiety, hunger or thirst\cite{fadok2018new}. Also, neuromodulators enable cortical circuits to process specific stimuli differentially and modify synaptic strengths in order to maintain short- or long-term memory traces of significant perceptual events and behavioural episodes. One of the major subcortical neuromodulatory systems for attention and arousal is the noradrenergic Locus Coeruleus(LC-NA)\cite{glennon2019locus}. Activity in the LC-GABA neurons controls the arousal by either activating or inhibiting LC-NA neurons\cite{breton2019active}. By preferentially targeting LC-GABA neurons, non-coincident inputs set thresholds for NA activation and enable modulation of tonic LC activity during different contexts\cite{breton2019active}. The activity in LC-NA neurons in turn regulate attention\cite{sara2012orienting, glennon2019locus, rodenkirch2019locus}, feature selectivity\cite{sara2012orienting, rodenkirch2019locus} and salience processing across the cortical\cite{vazey2018phasic} and thalamic nuclei\cite{benarroch2009locus, rodenkirch2019locus}. LC gets input from regions including CeA, PFC, hypothalamus, vagus nerve\cite{sara2012orienting} and projects to hippocampus, cingulate cortex, sensory cortices, somatosensory,  cerebellum\cite{schwarz2015viral}, thalamus\cite{benarroch2009locus, rodenkirch2019locus}, amygdala\cite{sara2012orienting} and prefrontal  cortex\cite{sara2012orienting}(fig~\ref{fig:model_}). The contextual modulation of LC is received via prefrontal regions and tunes the LC activity as per the environmental and cognitive contexts\cite{sara2012orienting}. For instance, LC response to a distractor, an unexpected event, is attenuated when the subject is focused on the task at hand, but the LC response to an awaited task-relevant cue is enhanced.

We have also observed the functional activity in the precuneus (BA7) in every calculated cognitive and affective functions. Precuneus as a cortical hub, has connections with many other cortical and subcortical parts of the brain\cite{cavanna2006precuneus}. Acquired autobiographical memories about events involve complex, multimodal and affectively salient memories embedded in a rich context of personal, social and environmental information. During regeneration, different level of spatio-temporal information is remapped. Based on our results, we suggest that precuneus is mediating connections among associative cortical and subcortical brain regions and playing an essential role in cognition-affect interaction. Although the functionality of precuneus has been explored, for example, in self-related processing, memory-related processing, attention, navigation and other tasks\cite{cavanna2006precuneus}, it demands more serious research attention. We emphasize its importance as it's widespread functional connections make it an integrating hub and elucidate its importance in emotion emergence. 

\subsection*{\textit{Cognition-Affect Interaction: The Anatomical Layout According to the Principles of Brain Organization}}
In the framework of statistical and hierarchical representation, the internally generated context is just a probabilistic activation of distributed pattern ~\cite{du2015pattern} throughout the cortex which is encoding the hidden causes of sensory experience or consequences~\cite{friston2010free}. The causal pattern of activation is embedded in the structural and functional asymmetry\cite{bastos2012canonical} in terms of the laminar projections and spatio-temporal hierarchy\cite{friston2018deep}. The hierarchical structure of information, which is distributed as a pattern, encapsulates abstract to event-specific sensory and emotional experience.

\begin{figure}[H]
\centering
\begin{subfigure}[t]{\textwidth}
\graphicspath{{pdf/}}
\includegraphics[width=0.9\textwidth, height=0.4\textwidth]{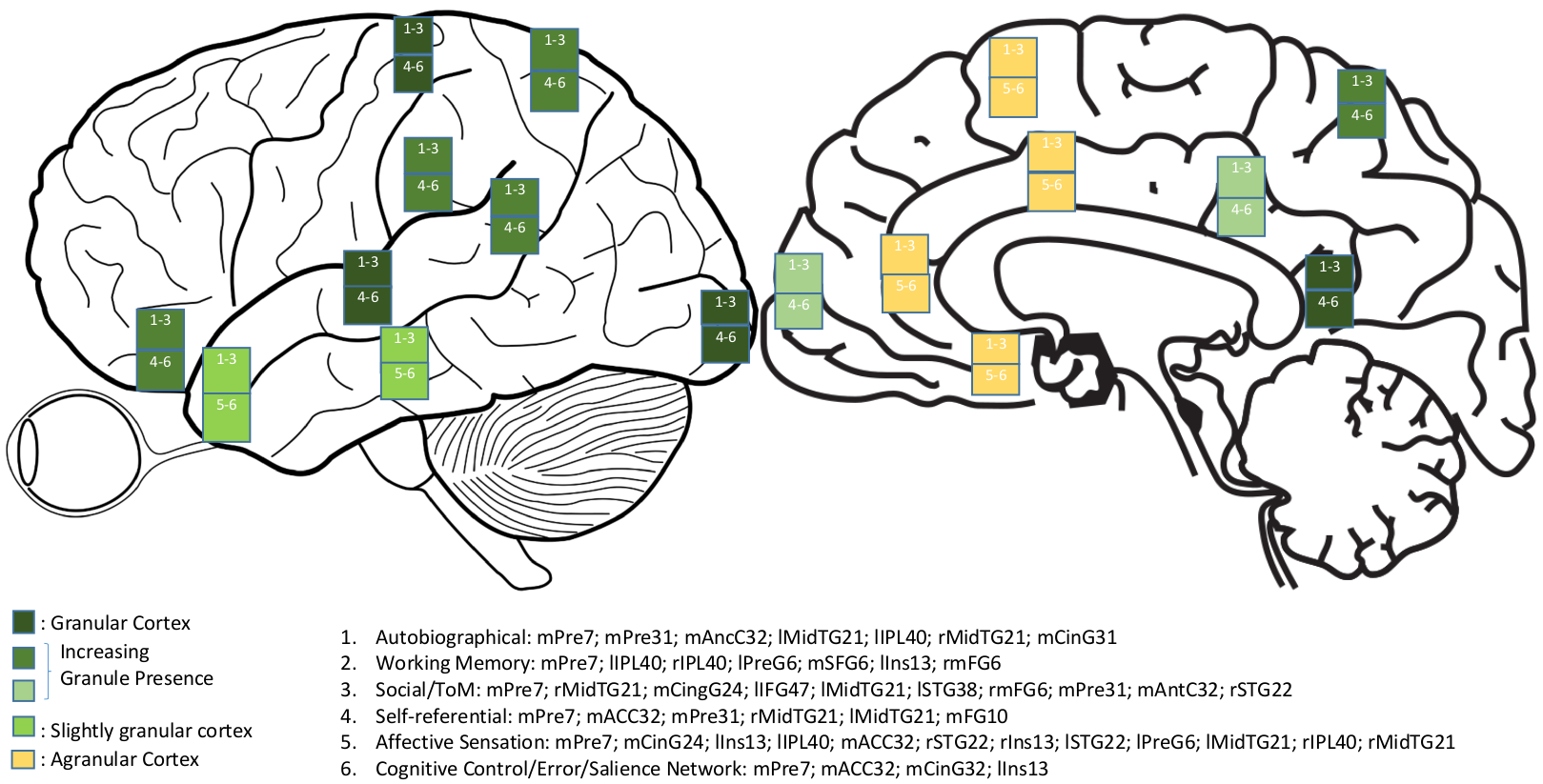}
\caption{}
\label{fig:Brain_Structure}
\end{subfigure}
\hfill
\begin{subfigure}[t]{0.48\textwidth}
\graphicspath{{pdf/}}
\includegraphics[width=0.8\textwidth, height=0.6\textwidth]{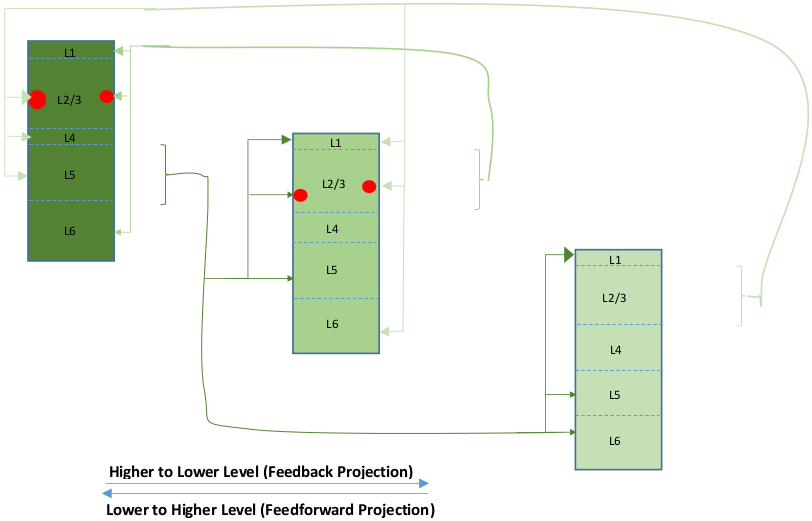}
\caption{}
\label{fig:Layer_wise_communication}
\end{subfigure}
\hfill
\begin{subfigure}[t]{0.48\textwidth}
\graphicspath{{pdf/}}
\includegraphics[width=0.8\textwidth, height=0.6\textwidth]{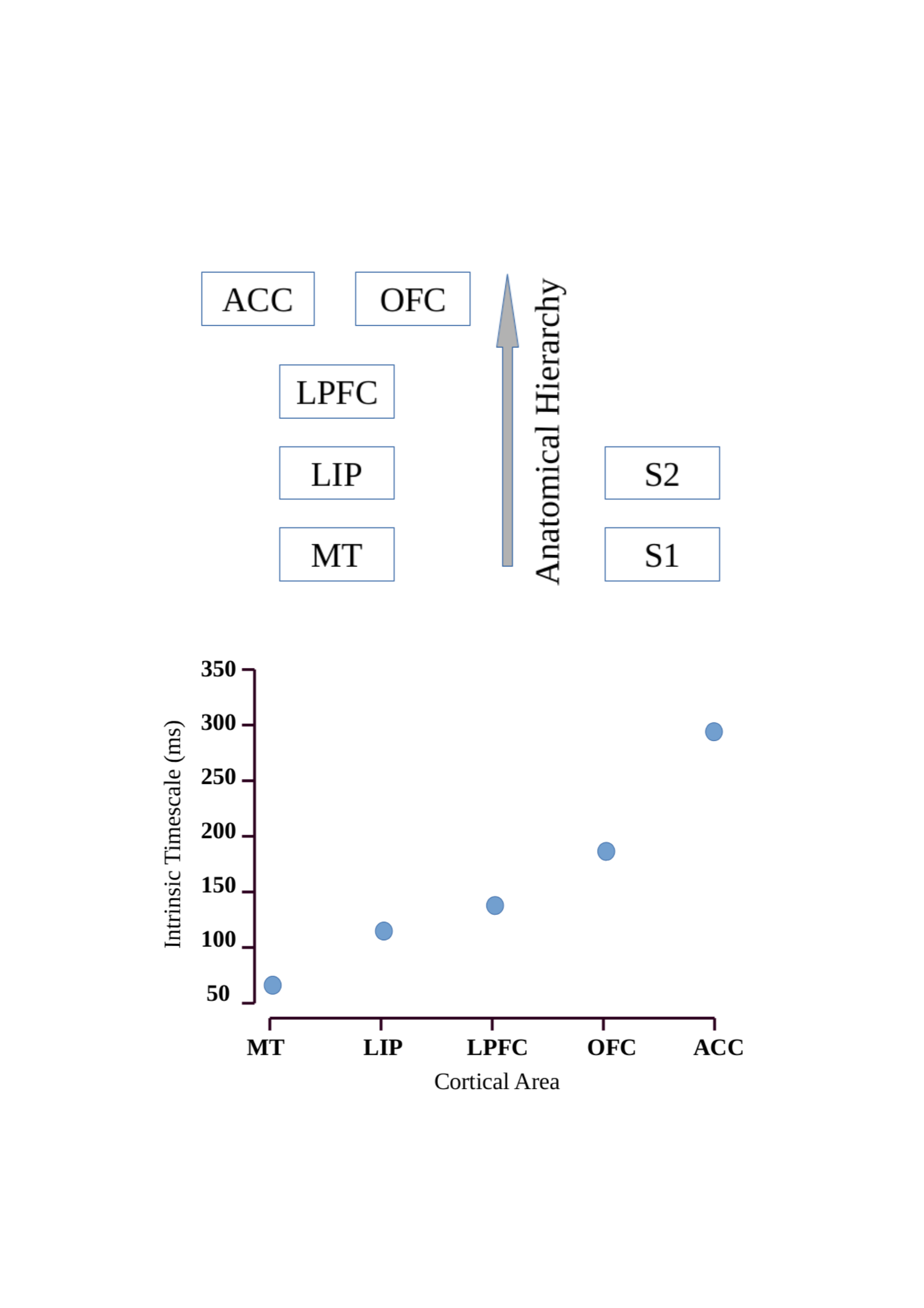}
\caption{}
\label{fig:HierarchyAnatomy}
\end{subfigure}
\caption{\textbf{Cortical structural and temporal functional organization underlying feedforward and feedback information integration in the cognition-affect integrated model of emotion.} (a)\textbf{Mesoscale cortical representation of active regions with granular structure:} Hierarchical agranular, increasing granular to fully expressed granular structures and anatomical projections adapted from\cite{von2009cellular, barbas1997cortical} and the depicted cortical regions are from our results; (b) \textbf{Interlayer Communication:} The diagram depicts interlayer communication among granular and slightly granular layers. The interlayer communication is between nearby layers and remote layers\cite{d2016recruitment}; The projections of feedback pathways diffuse among layer-1, 2/3 and 5 for nearby regions which progressively corner towards layer-1 and layer-6 for remote regions to modulate the centre-surround pattern and converge the cortical sub-cortical loops. On the contrary, the projections of feedforward pathways are diffused among layer-1, 2/3 and 6 for nearby regions which progressively projects to interior layers-2/3/4, and 5 for remote regions to perform the stimulus-driven activity. (c) \textbf{Structural and Temporal Hierarchy of Cortical Processing: } Structural hierarchy is mostly the result of degree of presence of granular layer in the region whereas temporal hierarchy is associated with the invariance property with higher-level cortical regions are showing higher invariance (less dynamic) in comparison to lower level which encodes more dynamic, less invariant and implicit representations. Adapted with permission from\cite{murray2014hierarchy}(Permission from nature neuroscience to reproduce the figures from\cite{murray2014hierarchy}). The structural and temporal hierarchical representation makes the bases of hierarchical context representation with more time-variant details in granular layers and less time-variant details in the agranular layers.}
\label{fig:HierarchicalStructure}
\end{figure}

\begin{figure}[H]
\begin{subfigure}[t]{0.5\textwidth}
\centering
\graphicspath{{pdf/}}
\includegraphics[width=\textwidth, height=0.8\textwidth]{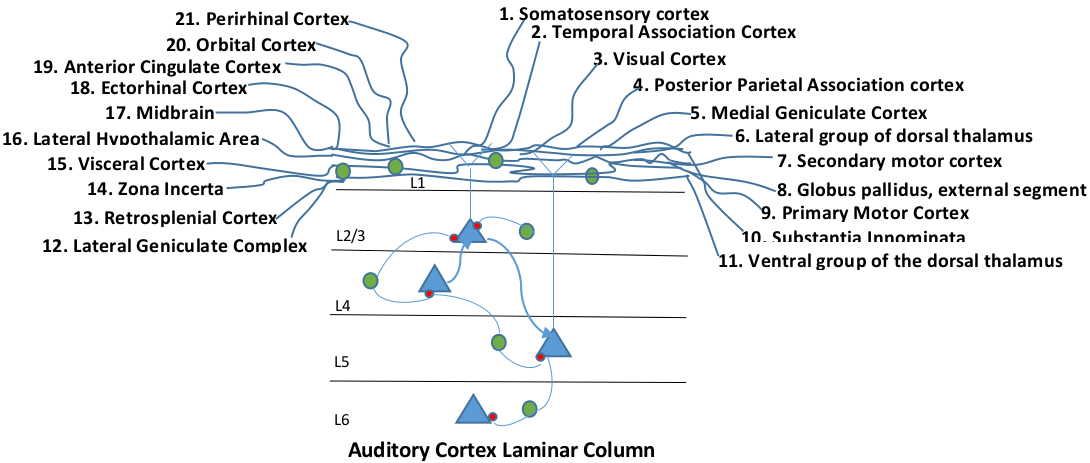}
\caption{} 
\label{fig:Layer_1_Projections}
\end{subfigure}
\hfill
\begin{subfigure}[t]{0.5\textwidth}
\centering
\graphicspath{{pdf/}}
\includegraphics[width=\textwidth, height=0.8\textwidth]{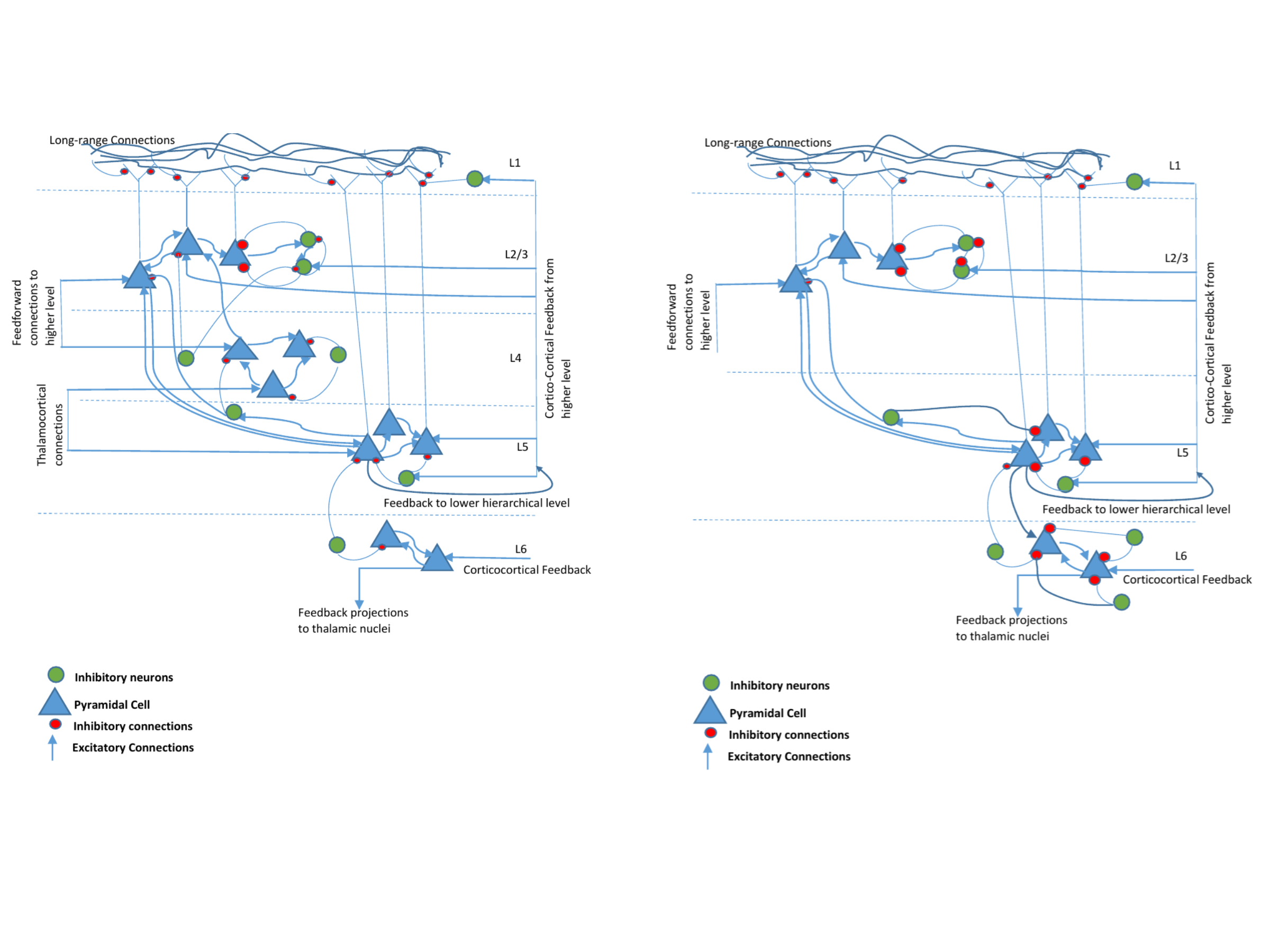}
\caption{} 
\label{fig:Granular_Agranular}
\end{subfigure}
\caption{\textbf{The cognition-affect integrated model of emotion: microcircuit level integration.}(a) \textbf{Long-range projections to layer-1:} The data is adapted from\cite{abs2018learning}. With the retrograde tracing, all the sources of axons in layer-1 of the auditory cortex is traced. It shows that inputs from regions with diverse functionalities projects to layer-1 of the cortical laminar column. The projections are carrying both cortico-cortical and cortico-subcortical signals. Long-range signals play an important role in modulating the functional microcircuits and centre-surround pattern in the target columns;(b)\textbf{Intralaminar organization of granular and agranular layers:} The highly differentiated granular columns contain granule cells and receive thalamic sensory inputs whereas agranular columns receive driving input in layer-2/3. Layer-1 is receiving modulatory long-range projections. The feedforward signals project to layer-2/3/4, whereas feedback signals project to all the layers except layer-4\cite{snyder2016dynamics, yang2013distinct, fino2011dense, weiler2008top, katzel2011columnar}. The recurrent circuit including inhibitory and excitatory neurons create neural ensembles organized in the centre-surround patterns to encode information.}
\label{fig:Microstructure}
\end{figure}

\textit{Hierarchical structure and feedforward-feedback projections: } \\
In the fig~\ref{fig:Brain_Structure} the sketch of the brain is depicted with the colour-coded rectangles illustrating the cellular organization of the regions\cite{von2009cellular} (these regions we found in our study). Agranular(see right side of fig~\ref{fig:Granular_Agranular}) and slightly granular layer is lacking and has negligibly developed granular layer 4, respectively. The granular layer 4 (left side of fig~\ref{fig:Granular_Agranular}) contains fine granule cells which are located in the primary sensory cortices. High-frequency burst activity in granule induces short-lived facilitation to ensure signalling within the first few spikes, which is rapidly followed by a reduction in the neurotransmitter release\cite{van2013high}. The fast rate coding of granule cells may be facilitating the sensory cortex to integrate changes in input at the faster rate and transmitting it to complex cells after performing spatiotemporal filtering\cite{van2013high}.  

A general information propagation scheme among the granular and agranular columns follows the distance rule(see fig~\ref{fig:Granular_Agranular}). Information from the sensory cortex which has fully developed layers follow the pathways through decreasing granularity to agranular cortices(feedforward pathway). On the contrary, information from the agranular cortex(majorly limbic cortex) can fan out information to increasing granular cortical regions progressively and finally terminate on the sensory granular layer(feedback pathway). The projections of feedback pathways diffuse among layer-1, 2/3 and 5 for nearby regions which progressively corner towards layer-1 and layer-6(fig~\ref{fig:Layer_wise_communication}) and modulate the representation by influencing centre-surround pattern. Corticothalamic nuclei in layer-6 projects back to thalamic nuclei. These projections create a cortico-thalamic loop (fig~\ref{fig:Granular_Agranular}) which contributes significantly in integrating higher-order cognitive functions. 

On the contrary, the projections of feedforward pathways are diffused among layer-1, 2/3 and 6 for nearby regions which progressively projects to interior layers-2/3/4, and 5(fig~\ref{fig:Layer_wise_communication}) to project stimulus-driven pattern in the centre-surround configuration. Starting from the sensory cortex, which is having the fully expressed granular layer 4, the driving feedforward signal (modulated by feedback pathways at every stage) follows the structural descendants in terms of granularity and towards superficial layer (layer-2/3) and inferior layer (layer-5) of agranular cortices to limbic system\cite{song2019crucial}. These feedforward and feedback pathways create cortico-cortical and cortico-subcortical loops, for example, cortico-striato-thalmic loop forms segregated sub-cortical loops which integrates cognitive and affective aspect of behaviour\cite{metzger2010high}.

\textit{functional microcircuits and neural ensembles:} \\
Asymmetry in anatomical projection also favours the context and cognitive modulation of lower-order sensations. For both the feedforward and feedback projections, with the distance, the I/E asymmetry is more than if two regions had been in the nearby locations. Moreover, it is more asymmetric for the feedforward projections than for the feedback projections\cite{d2016recruitment}(fig~\ref{fig:Granular_Agranular}). A large proportion of inhibitory activity due to long-range projections in L-2/3 is regulating the encoding of information by performing inhibition in the neural ensemble of L-2/3 and thus might be encouraging the biasing for inference based internal representations of concepts and contexts. In the feedforward direction, due to increasing inhibition ratio with the longer distance, in higher regions, the excitatory activity (due to sensory input) is less than the inhibitory activity which might help codify complex abstract representations with less in lower-level details than in the nearby sensory regions\cite{ardid2007integrated, zhang2014long, kuchibhotla2018neural, lee2013disinhibitory, letzkus2015disinhibition, artinian2018disinhibition}. 

The long-range projections from higher regions modulate the functional microcircuits of lower-level regions and vice-versa. In the laminar organization of cortical columns, the recurrent circuits (fig~\ref{fig:Granular_Agranular}) involve both excitatory and inhibitory neurons which, in general, creates the centre-surround pattern. These recurrent circuits\cite{seeman2018sparse} are driven and modulated by bottom-up and top-down projections, respectively. Top-down projections enhance the firing rates of putative inhibitory inter-neurons \cite{zhang2014long}. These different types of inhibitory interneurons play different roles in the top-down modulation. For example, in response to focal Cg axon activation, SOM+ and PV+ neurons inhibit pyramidal neurons over a broad cortical area (with SOM+ neurons as a major source of surround inhibition at 200 $\mu m$), whereas VIP+ neurons selectively enhance the responses at 0 $\mu m$ by localized inhibition of SOM+ neurons\cite{pfeffer2014inhibitory, karnani2016opening}(facilitating centre by dis-inhibiting selective pyramidal cells). The long-range projections in general target VIP+ neurons than other neuron types\cite{pi2013cortical} in layer 2/3. The disinhibitory effect of VIP+ neurons on pyramidal neurons is reported in somatosensory\cite{wood2017cortical, lee2013disinhibitory}, visual\cite{wood2017cortical, wilson2014visual, mesik2015functional}, auditory\cite{wood2017cortical, pi2013cortical, mesik2015functional, kuchibhotla2018neural}, mPFC\cite{pi2013cortical}, cross-modality sensory projections\cite{ibrahim2016cross}, and learning and memory\cite{letzkus2015disinhibition, francavilla2015coordination, artinian2018disinhibition}. This centre facilitation and surround suppression mechanism by top-down modulation are equivalent to the bottom-up centre-surround mechanism. Moreover, projections from individual neurons of remote regions(for example, Cg\cite{zhang2014long}) selectively projects to restricted neurons of the target regions(for example, top-down projections allow targeted spatial modulation in V1\cite{zhang2014long}). The gain effect to this attentional activity, according to "feature-similarity gain modulation principle", depends on similarity and difference between the attended feature (bottom-up activity) and the preferred feature (top-down activity) of the neural population\cite{ardid2007integrated}. In the case of similarity, the centre-surround effect due to both top-down and bottom-up activity match and scale the tuning curve in a multiplicative manner. On the contrary, in the case of mismatch of centre-surround effect, suppression of the tuning curve takes place\cite{zhang2014long}. Other than the cortico-cortical projections, limbic cortices issue widespread projections from their deep layers and reach eulaminate areas by terminating in layer-1\cite{barbas1995anatomic} and thus modulate cortical representations.

Long-range anatomical projections, feedforward and feedback layers specific targeted projections, cortico-subcortical loop, neural ensemble in the form of centre-surround pattern and its modulation due to global state of the brain, represents the functional organization of the brain in the structural or anatomical frame. The information, encoded in topological(local functional circuits) and distributed global neural patterns, is communicated via anatomical neural projections to give bases for cognition-affect interaction.

\section*{Concluding Remarks and Future Perspectives}

Based on our analysis and results, we infer that emotions are the product of the interaction of cognition and affect in a loop. Emotion can not be attributed to a definite marker, but they are dynamically decided statistical and fuzzy responses in the frame of the contexts existing at that moment. Emotion processing utilizes large-scale brain networks\cite{barrett2013large, pessoa2017network} which are related to cognitive functions, for example, salience network, autobiographical events(episodic memory) related system and social processing related system\cite{jakobs1997emotional, sterling2012allostasis}. So, emotions are constructs or concepts which aim to avoid danger and approach social gain in order to achieve learned allostatic stability. Emotions follow structural and temporal hierarchical organization ~\cite{parr2017working}. Varieties of activities which are happening at the lower sensory and autonomic control level can belong to the same emotion category, and different emotions can share some of the common features at the sensory and lower level~\cite{barrett2012emotions}.

\definecolor{grisframe}{gray}{0.95}
\definecolor{gristitleframe}{gray}{0.85}
\mdfsetup{backgroundcolor=grisframe,
 skipabove=5pt,
 skipbelow=5pt,
 leftmargin=0pt,
 rightmargin=0pt,
 innertopmargin=6pt,
 innerbottommargin=6pt,
 innerleftmargin=16pt,
 innerrightmargin=6pt,
 frametitleaboveskip=6pt,
 frametitlebelowskip=6pt,
 frametitlerule=true,
 frametitlebackgroundcolor=gristitleframe}

\begin{mdframed}[frametitle={Outstanding Questions}]
\begin{itemize}
    \item Where and up to what extent does the classical notion of basic emotions and their universality fits in this model? In other words, how much context-dependent are “basic” emotions?
    \item What role the structural hierarchy plays in the relevancy and quality of emotional response?
    \item What role the hierarchy of temporal processing in cortical regions plays in emotional response time? 
    \item What is the cognition-affect interaction dynamics which can differentiate the emotional event from the non-emotional event at the functional connection-level?
    \item What is the contribution of integration hubs and their interactions in cognition-affect modulation and evolution of emotion?
\end{itemize}
\end{mdframed}

\definecolor{grisframe}{gray}{0.95}
\definecolor{gristitleframe}{gray}{0.85}
\mdfsetup{backgroundcolor=grisframe,
 skipabove=6pt,
 skipbelow=6pt,
 leftmargin=0pt,
 rightmargin=0pt,
 innertopmargin=6pt,
 innerbottommargin=6pt,
 innerleftmargin=6pt,
 innerrightmargin=6pt,
 frametitleaboveskip=6pt,
 frametitlebelowskip=6pt,
 frametitlerule=true,
 frametitlebackgroundcolor=gristitleframe}

\begin{mdframed}[frametitle={Glossary}]
\textbf{Allostasis:} Brain anticipates needs and provides organism's physiological infrastructure to fulfil these needs in order to regulate the organism's internal milieu\cite{sterling2015principles}. Allostasis is different from homeostasis in that it has experience based variable stable point regulating the organism's behavior, whereas homeostasis is based on constant setpoint and an error mechanism to regain this constant set point\cite{sterling2012allostasis}. \\
\textbf{Core Affect:} Core affects are the behavioural-action tendencies with instinctual-arousal tools of nature rather than constructions of nature. Core affects reflect relatively invisible neurodynamics of ancient brain systems. In the words of Jaak Panksepp "at their core, raw affective experiences appear to be pre-propositional gifts of nature—cognitively impenetrable tools for living that inform us about the states of our body, the sensory aspects of the world that support or detract from our survival, and various distinct types of emotional arousal that can inundate our minds. Affects reflect the heuristic value codes that magnificently assist survival, and give ‘value’ to life." \\
\textbf{Predictive coding:} A finding by\cite{rao1999predictive} which signifies top-down and bottom-up processing as feedback and feedforward projections carrying prediction based on the inference from past and prediction error(in terms of predicted minus what is actually observed), respectively. \\
\textbf{Recurrent Circuits or Microcircuits or Neural ensemble:} A group of excitatory and inhibitory cells which co-activate together in a specific pattern upon receiving a cue and performing specific information processing relevant to a task. For example, neural ensemble creating concepts in the hippocampus, orientation columns in visual cortex and so on.
\end{mdframed}

\section*{Methods}

\subsection*{Participants:} Deap Data, which is freely available online, is used\cite{koelstra2011deap}. Thirty-two healthy participants in two separate locations Twente (22 participants) and Geneva (10 participants) participated in the study. All of them singed an informed consent form before starting\cite{Deap2020Data, koelstra2011deap}. The study sample comprised of right-handed 17 males and 15 females aged between 19 and 37 (mean age $27.19\pm4.44$; right-handed; undergraduate and postgraduate; normal or corrected to normal vision; no history of neurological, psychiatric diseases or substance-related disorders; no significant general medical condition). Before the experiment, each participant signed a consent form and filled out a questionnaire.  

\subsection*{Stimulations:} All the experimental stimuli were carefully filtered down from a collection of 120 stimulus videos to the final 40 test video clips chosen by using web-based emotion assessment interface. Stimuli and rating scales were presented using the software by Neurobehavioral systems on a 17-inch screen (1280x1024, 60Hz) with the 800x600 resolution to minimize eye movements. Participants were sitting approximately 1 meter away from the screen.

Music videos were used to elicit emotions in the participants during the experiment. Anticipation is the key to understand, comprehend and feel emotions while listening to music. Musical anticipation itself can evoke a variety of emotions~\cite{vuust2008anticipation}.
The music-evoked emotions are comprised of three principles\cite{koelsch2014brain}. The first principle is serving the social function, second is related to musical expectancy and tension (due to the harmonic structure of music), and third is related to emotional contagion. The social function of emotional music is classified further in functions related to social cognition(understanding composer's intention), co-pathy(empathically affected emotional homogeneity), social and emotional regulation through communication, action coordination and group cooperation. . 
Music is social; we inherently feel the social value of reaching to others in music or by moving others in a song across the broad social milieu~\cite{schulkin2014evolution, wallmark2018neurophysiological}. 

\subsection*{Experimental Protocol:}
The raw data, information about funding resources and ethical approval is available on\cite{Deap2020Data, koelstra2011deap}.      
All the participants were given a set of instructions about the experiment protocol and explained the meaning of the different scales used for self-assessment. A practice trial for each participant is conducted to familiarize participants with the experiment. After the practice trial, the experimenter left the room and the participant started the experiment with a keypress on a keyboard. 

Initially, a baseline recording of 2 minutes while participants were looking at the fixation cross is done. It followed by the presentation of 40 trial video in the following paradigm:
\begin{enumerate}
    \item To inform the participants about the current trial a 2-second screen displaying trial no,
    \item baseline recording with the display of fixation cross for 5 seconds,
    \item display of trial for 60 seconds,
    \item Rating scale of valence, arousal, dominance, familiarity, and liking for the self-assessment.     
\end{enumerate}

A short break after 20 trials were given to the participants and participants were offered with some cookies and non-alcoholic/non-caffeinated beverages during this time. After the break remaining trials followed above-mentioned steps.

\subsection*{Scalp Recording:} The experiment was performed in two laboratory environments with controlled illumination. EEG and peripheral physiological signals were recorded using a Biosemi ActiveTwo System. In experiment \cite{koelstra2011deap}, 32 EEG active AgCl electrodes(10-20 system) and 8 peripheral physiological channels are used to record brain activity and peripheral physiological signals, respectively, while subjects were watching 1-minute emotional video excerpts. Each subject watched 40 excerpts. EEG was recorded at a sampling rate of 512 Hz.

\subsection*{EEG pre-processing to functional connectome}
The methodology is depicted in fig~\ref{fig:Methodology}. Using bioSig Matlab toolbox and EEGLAB the unprocessed DEAP data \cite{koelstra2011deap} is extracted. For preprocessing, the Makoto's preprocessing pipeline is followed. From the continuous stream of data, EEG signals for emotion and baseline is extracted from the raw data. Originally data is recorded at 512Hz which is down-sampled to 128Hz. Using high pass filter at cut-off frequency 4.0Hz data is filtered. Again, using a low pass filter with cut-off frequency 45.0Hz data is filtered. We didn't find any bad channels in the data. Data is re-referenced to average. ICA is applied to detect artefacts in the signal due to eye blink and major muscle movements (see figure-S1 in the supplementary section). 

Narrowband theta oscillations (4-8Hz) have been considered for analysis. Reason behind selecting this band is that previous studies found synchronization in theta band for processing emotional modalities \cite{chen2015integration,jiang2017event, bocharov2017depression,han2016enhanced,knyazev2010gender,del2016voice,symons2016functional, knyazev2009event}. Moreover, reduced synchronized activity in the theta band has been reported in case of emotional disorders \cite{bocharov2017depression,knyazev2015predisposition,csukly2014event, aftanas2003disruption}. Also, there is plenty of evidence that during the mental reconstruction and holding of the information in working memory theta phase synchronization takes place.

\subsubsection*{EEG Geosource Localization: } 
The process of source localization involves forward and inverse modelling. Calculation of scalp potentials from the current sources in the brain with the help of some physical theory is said to be modelization or simulation or forward problem. Given the electrode potentials recorded at the distinct brain scalp sites, geometry and conductivity within the brain, estimating the location and magnitude of the current sources responsible for generating these potentials is the EEG inverse problem. The EEG inverse problem is an ill-posed problem as $N_V >>> N_E$ \cite{song2015eeg}. This ambiguity is constrained using the number of sources, spatial smoothness, spatial sparsity, and the combination of sparsity, as well as constraints on the dynamics of the source time courses \cite{mahjoory2017consistency}. Source localization problem had been the focus of interest for modelling community for decades and approached with different solutions: minimum norm, LORETA, sLORETA, eLORETA, MUSIC, FOCUSS, and ICA (see survey \cite{jatoi2014survey}). 

In this study we have used standardized low resolution electromagnetic tomography (sLORETA) method \cite{pascual2002standardized}. It is a distributed inverse imaging method. The current density estimate is based on the minimum norm ($l_2 norm$) solution, and localization inference is based on standardized values of the current density estimates. sLORETA is capable of exact (zero-error) localization. The objective function to be minimized to get zero error localization is

$F = \| \Phi - \textbf{KJ} - c1 \|^2 + \alpha \|\textbf{J}\|^2$

where $\alpha \geq 0$ is a regularization parameter. This functional is to be minimized with respect to \textbf{J} and c, for given \textbf{K}, $\Phi$ and $\alpha$. The explicit solution to this minimization problem is

$\hat{\textbf{J}} = \textbf{T}\Phi$

where: 

$\textbf{T} = \textbf{K}^T\textbf{H}[HKK^TH + \alpha H]^+$

$\textbf{H} = \textbf{I} - \textbf{11}^T/\textbf{1}^T\textbf{1}$

with $\textbf{H} \in \mathbb{R}^{N_E * N_E}$ denoting the centering matrix; $\textbf{I} \in \mathbb{R}^{N_E * N_E}$ the identity matrix; and $\textbf{1} \in \mathbb{R}^{N_E x N_E}$ is a vector of ones.

MNE library \cite{gramfort2013meg} is used to perform source localization, and visualization of the activity is done using nilearn python library \cite{abraham2014machine, huntenburg2017loading}.

\subsubsection*{Networks Analysis:}
Application of network science in studying the connectome of the brain is revealing more significant functional insights of it. Anatomical and functional connectome of the human brain helped in parcellating brain structure at a refined level~\cite{glasser2016multi}. Voxel to voxel connectivity is calculated using PLV. Since in total the data size to process was extensive and time-consuming, we implemented the correlation connectivity using PLV values \cite{bruna2018phase} in GP-GPU. PLV value is calculated by transforming the real signal into the analytical signal using the Hilbert transform. For the correlation-based functional connection analysis, PLV value is calculated with the following formulation. 

$PLV_{ij}(t) = \frac{1}{N} |\Sigma_{n=1}^{N}e^{-i(\psi_i{(t,n)}-\psi_j{(t,n)})}|$

This calculation resulted in a matrix of 6239x6239(all voxel connectivity). 

\subsubsection*{Voxel-pairwise functional decoding: }
Each voxel pair calculated from our PLV based network analysis and filtered for significance based on the permutation test with $p-value < 0.0002$ is selected. Communities were calculated for the obtained functional networks(explained in the next subheading). These voxel pairs were decoded for their associated functions using Neurosynth meta-analysis database. It is achieved in two steps: first, a co-activation network with 0.1 as activation threshold(threshold for a study to be included based on amount of activation displayed) and 5mm as radius(the distance cut-off for inclusion of studies with percentage of activation of seed pairs) and second, decoding the functions and their meta-analytic co-activation associated with the activity in the supplied voxel pair. The first step gave us the network-based co-activation map for the provided voxel pair as a seed and the second step provided us with a correlational factor signifying the association of the function with the probed voxel pair. Out of a list of functions, we have selected only those functions for which the correlation value were more than 0.3 and discarded other functions. Although there is no standard for these thresholds, the decision on the threshold was taken based on the prior knowledge about 'can this voxel pair possibly be related to the particular set of functions'. Likewise, for all the voxel pairs, the functions have been decoded. Only the functions with higher frequency were finally selected for further analysis. 
\subsubsection*{Finding Communities:}
Communities are the properties of real networks which could be characterized with comparatively dense intra-group connectivity than inter-group connectivity. The evidence of community structures in the brain signifies its segregation property, whereas the connection between these communities signifies the integration of these segregated functionalities of the brain. Validation of communities in real networks is done by comparison with benchmark graphs with a known number of communities and its size. In the study, LFR benchmark~\cite{lancichinetti2009benchmarks} is used to compare five different community detection algorithms. These are infomap~\cite{rosvall2008maps}, leading eigenvector, label propogation, multilevel, and edge betweenness~\cite{girvan2002community}. Among these communities, we found that infomap random walk algorithm is working superior to the other algorithms. Infomap random walk is reported to be better in finding communities than other algorithms~\cite{barabasi2016network, yang2016comparative}. 
We didn't find the correspondence between all the nodes in one community corresponding to any particular network out of many networks(for example, dmn, dorsal and ventral AN, SN, and so on). The reason behind this may be; first, the contribution of individual node in the network can be dissociated and can perform task-specific functionality, second, as per the task at hand, subclusters of nodes from different proposed networks(such as DMN, AM, AN, SN, social and theory of mind) can come together to form a large-scale task-specific integrated network. The community-wise connectome is shown in supplementary fig-S3.

\subsection*{MVPA Analysis using Deep Learning: }
Generalized training on four datasets(Mental arithmetic \cite{zyma2019electroencephalograms}, motor movement and imagery\cite{schalk2004bci2000}, grid of characters\cite{citi2010documenting}, and SSVEP EEG database\cite{oikonomou2016comparative}) is done on the convolution neural network(CNN), a deep learning architecture, (fig~\ref{fig:Deep_Model}) with the following specifications: four 2d convolution layer with 32 kernels and two 2d convolution layer with 64 kernels of size 3x3; relu activation function for the convolution layers and soft-plus activation function for output layer; adam optimizer with learning rate:0.0001(zero decay), $\beta _1$:0.9, $\beta _2$:0.999; categorical cross-entropy comparison for error calculation between actual and predicted class. Keras Python deep learning library with TensorFlow as the backend has been used for creating the CNN model with the parameters mentioned above. The model is trained on Nvidia Tesla-V100-PCIE data centre GPU with 16 GB capacity with the input tensor of 448x1230x99. 

We created 200 batches of training input for each of the considered datasets (the batch mentioned here should not be confused with deep learning batch size which is typically used during training to deal with the computational and convergence time). That means, total of 800 batches of training input with varying number of samples. Machine trained on one batch was used in the next batch and so on. In this way, the trained machine was quite robust on detecting general features of electrophysiological data. The details about feature calculation, input size, s/w and h/w specifications are discussed in the methods section. 

The input tensor (for the emotion data we have analyzed) had 448 samples distributed as 14 emotion stimuli for 32 subjects. Since we have used above mentioned four datasets to train our model, we have created in total 800 input batches(200 input batches per input training dataset) with varying number of samples (as per the subjects and outputs of the training data). 1230 are the number of voxels. This number has kept constant across the different above mentioned training datasets since 615 pairwise connections between regions are calculated using statistical significance analysis. In the input tensor 448x1230x99, the last dimension is representing statistical features (including median, standard deviation, mean, maximum, range, minimum, skewness, variance and kurtosis values) calculated on 9 segments constructed from the 60-second signal. The same statistical features were calculated for the whole signal adding 9 extra features on 90 features calculated from the 9 segments (creating in total 99 feature). These statistical features are calculated due to the trade-off among the increasing number of weight parameters for the architecture complexity and limitation of machine capability to deal with the large input size of 448x32x7680(If the whole signal for 60 seconds with 128 sampling rate had been considered). The trained general EEG model is used for the final testing on original voxel time-series for the same batch size. The true-positive rate and false-positive rate for the ROC analysis is calculated using the following formula:
$TPR = \frac{TP}{TP+FN}$
and
$FPR = \frac{FP}{FP+TN}$
where TP stands for true positive, FN stands for false negative, FP stands for false positive, and TN stands for true negative. To understand how the quantification of ROC and AUC describes the quality of classifier in distinguishing different classes, please see fig-S10 in the supplementary section.

%
%

\section*{Processed Data and Code Availability}
All the code and processed data will be made available upon publication.

\section*{Ethics declarations}
This study was carried out in compliance with the DEAP data~\cite{koelstra2011deap} (available online) end user license agreement (available on ~\cite{Deap2020Eula}). All the ethical guidelines provided in the above mentioned license agreement form have been rigorously followed. Consent information for each participant is included in the participant questionnaire file (available on ~\cite{Deap2020Data}). The authors also declare no competing interests.

\section*{Author contributions statement}

S.M and U.S.T. both have developed the presented idea and model. S.M did the coding whereas S.M and U.S.T. both discussed and decided the computational procedure. The interpretation of the calculated results is done by both U.S.T and S.M.. The first draft of the manuscript is prepared by the S.M. U.S.T. and S.M. both refined the manuscript to the presentation and submission level. 

\section*{Additional information}
The corresponding author is responsible for submitting a \href{http://www.nature.com/srep/policies/index.html#competing}{competing interests statement} on behalf of all authors of the paper. 

\newpage
\section*{Figures}

\begin{figure}[H]
\centering
\begin{subfigure}[H]{\textwidth}
\centering
\graphicspath{{pictures/}}
\includegraphics[width=\textwidth, height=0.15\textwidth]{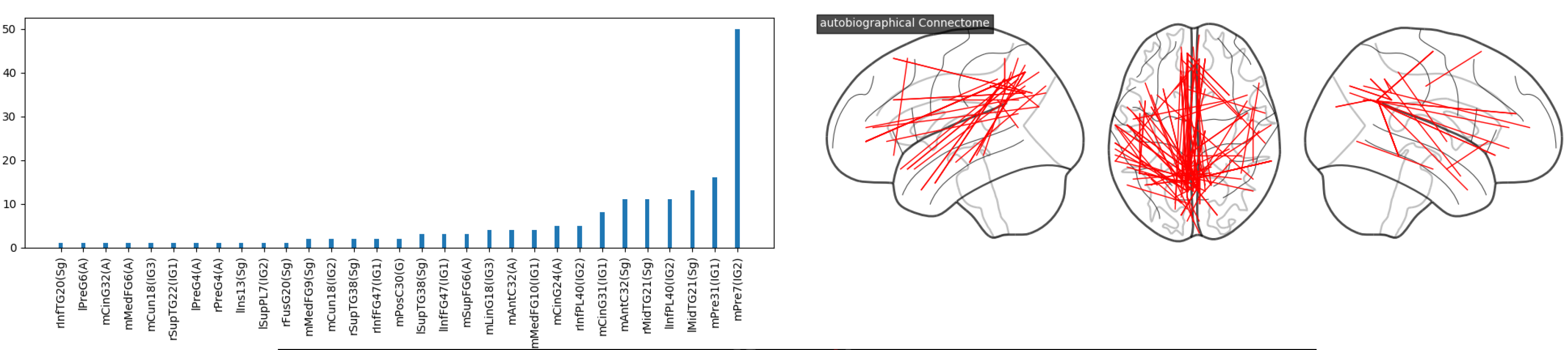}
\caption{{\tiny Autobiographical or Episodic Memory}} 
\label{fig:connect_auto}
\end{subfigure}

\begin{subfigure}[H]{\textwidth}
\centering
\graphicspath{{pictures/}}
\includegraphics[width=\textwidth, height=0.15\textwidth]{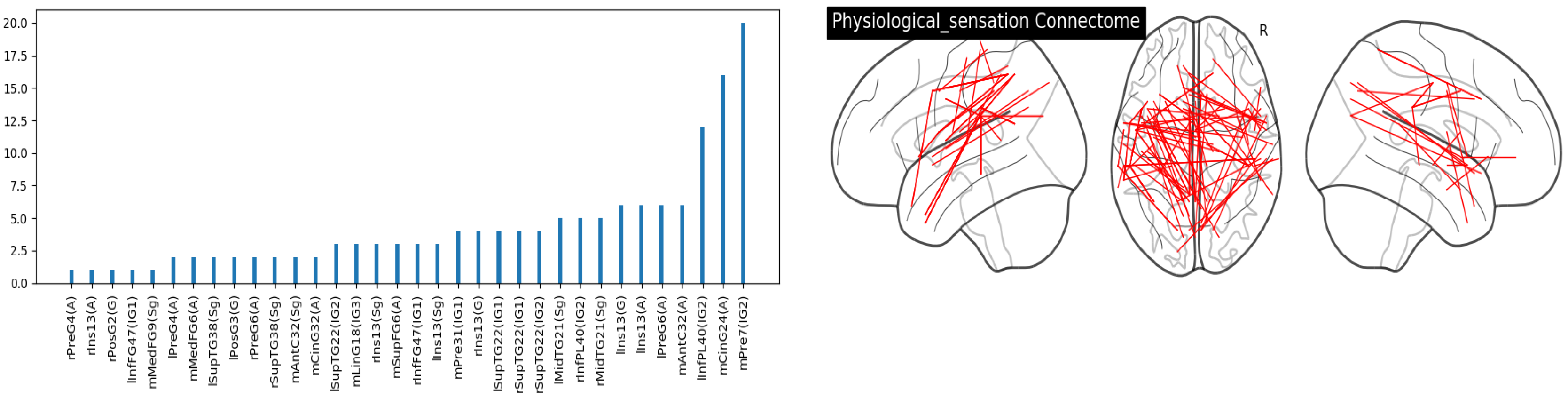}
\caption{{\tiny Affective sensation and Pain}} 
\label{fig:connect_pain}
\end{subfigure}

\begin{subfigure}[H]{\textwidth}
\centering
\graphicspath{{pictures/}}
\includegraphics[width=\textwidth, height=0.15\textwidth]{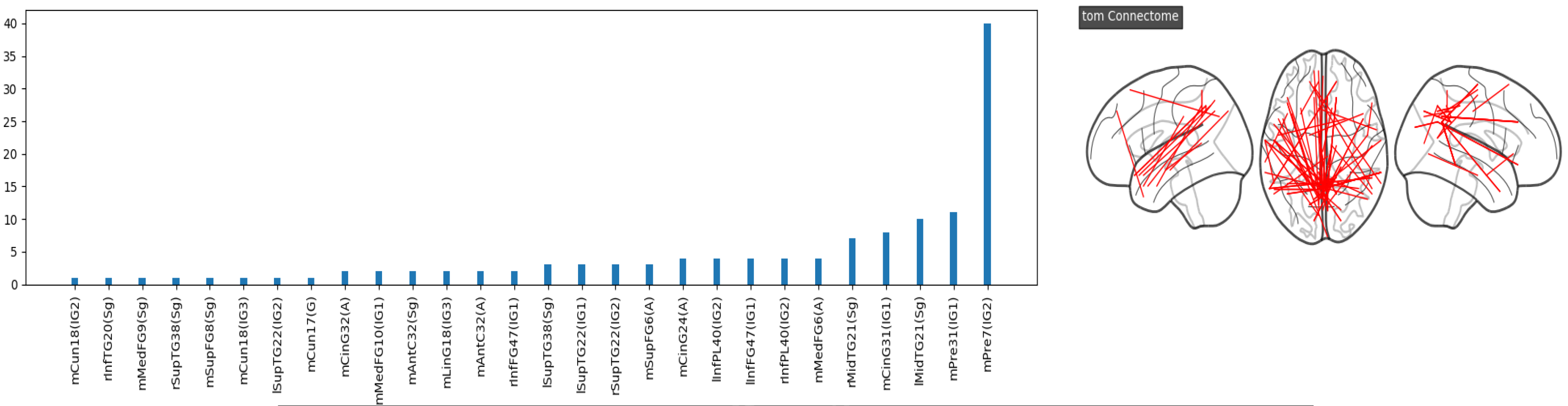}
\caption{{\tiny Theory of Mind}} 
\label{fig:connect_tom}
\end{subfigure}

\begin{subfigure}[H]{\textwidth}
\centering
\graphicspath{{pictures/}}
\includegraphics[width=\textwidth, height=0.15\textwidth]{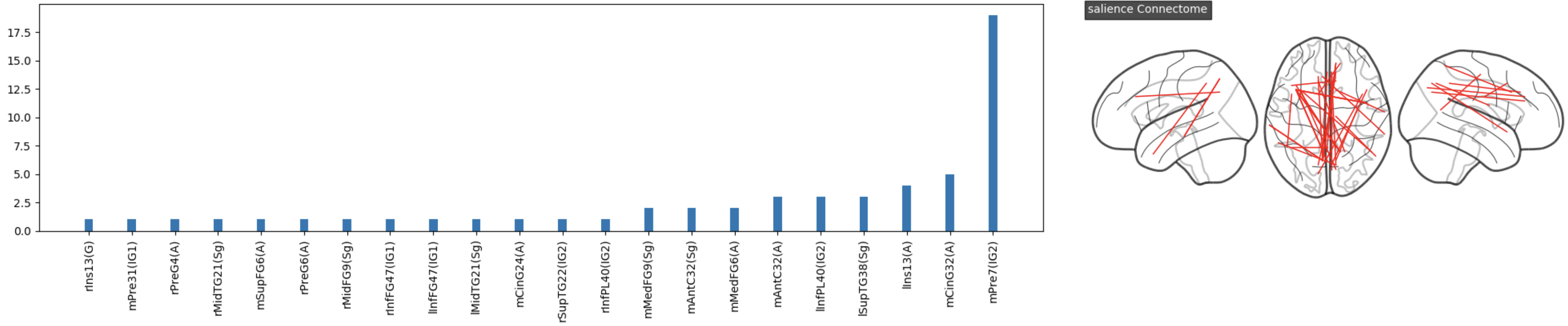}
\caption{{\tiny Salience Network}} 
\label{fig:connect_saliency}
\end{subfigure}
\end{figure}

\begin{figure}[H]\ContinuedFloat
\begin{subfigure}[H]{\textwidth}
\centering
\graphicspath{{pictures/}}
\includegraphics[width=\textwidth, height=0.1\textwidth]{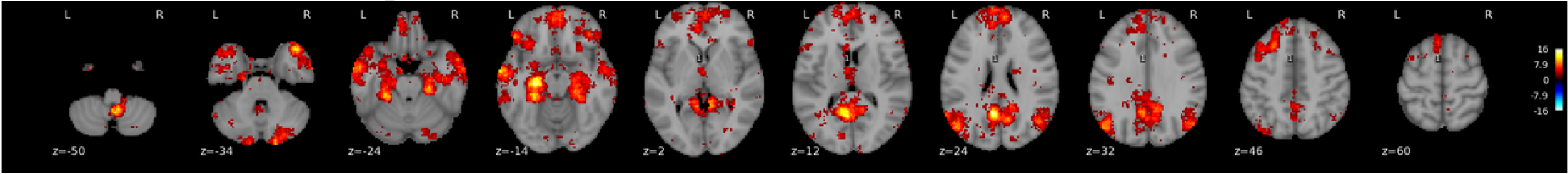}
\caption{{\tiny Autobiographical or Episodic Memory}} 
\label{fig:statmap_auto}
\end{subfigure}

\begin{subfigure}[H]{\textwidth}
\centering
\graphicspath{{pictures/}}
\includegraphics[width=\textwidth, height=0.1\textwidth]{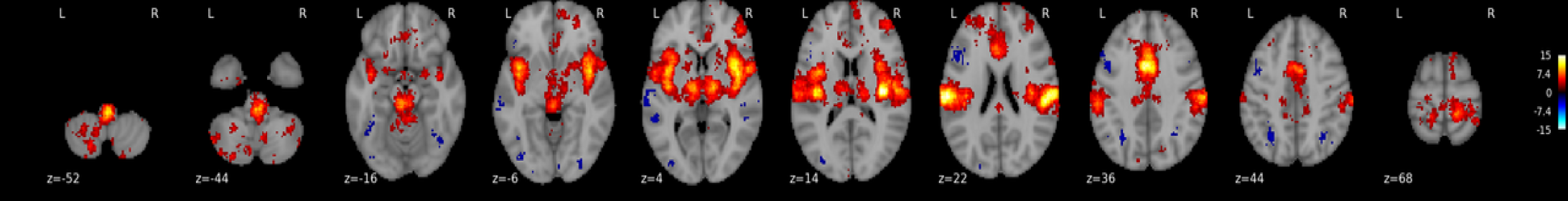}
\caption{{\tiny Affective sensation and Pain}} 
\label{fig:statmap_pain}
\end{subfigure}

\begin{subfigure}[H]{\textwidth}
\centering
\graphicspath{{pictures/}}
\includegraphics[width=\textwidth, height=0.1\textwidth]{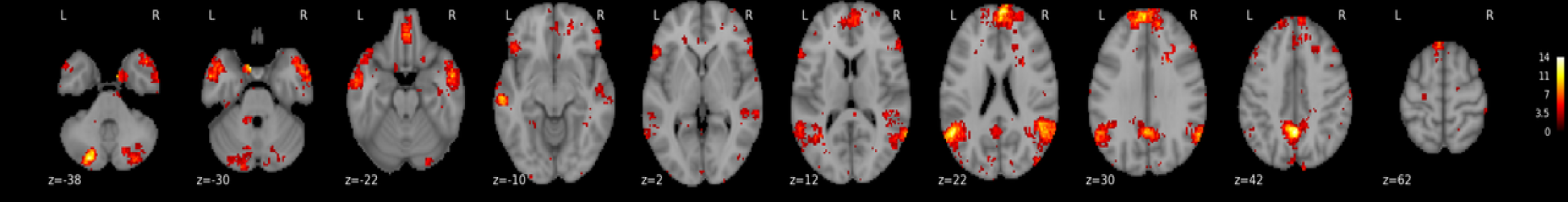}
\caption{{\tiny Theory of Mind}} 
\label{fig:statmap_tom}
\end{subfigure}

\begin{subfigure}[H]{\textwidth}
\centering
\graphicspath{{pictures/}}
\includegraphics[width=\textwidth, height=0.1\textwidth]{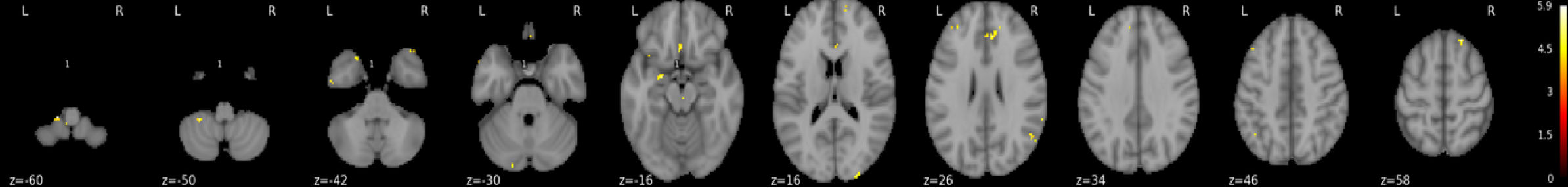}
\caption{{\tiny Salience Network}} 
\label{fig:statmap_saliency}
\end{subfigure}

\begin{subfigure}[H]{\textwidth}
\centering
\graphicspath{{pictures/}}
\includegraphics[width=\textwidth, height=0.2\textwidth]{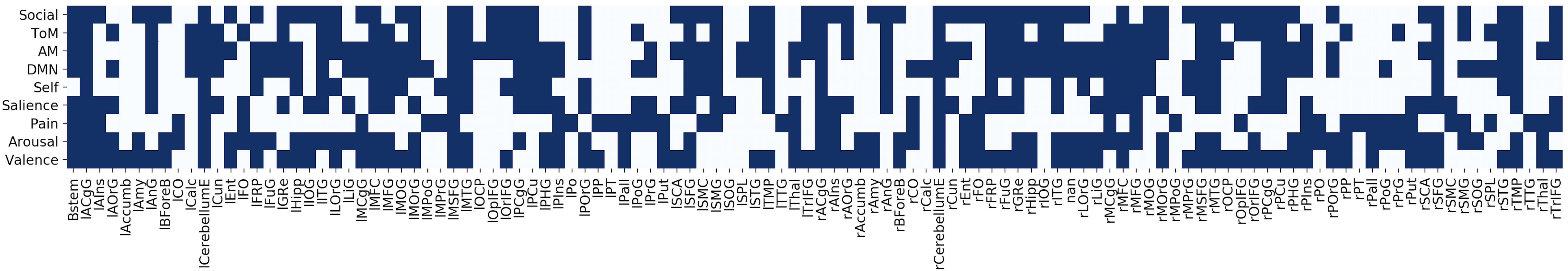}
\caption{{\tiny Heat Map}} 
\label{fig:regionheapmap}
\end{subfigure}

\caption{\textbf{Functionally phase-locked activity of brain regions and decoded cognitive functions:} ANOVA two-sided t-test, for the categorical subtraction between baseline functional connectivity and emotional task-related functional connectivity, is designed on 32 human subjects(female-15 and male-17) for 12 different emotion conditions to get significant phase-locked activity between different brain regions. Three emotions we discarded since they didn't pass our permutation test. Non-parametric permutation with FDR correction (the alpha level of 0.0002) is implemented, and significant connections were considered to form the final network. Weights of the link in the network are decided based on the number of voxel-pair connections between two regions. A high value indicates more correlated activity between two brain regions. Now those pair of regions which had connection probability more than 99.5 per cent, out of all voxel pair connections, had been taken. For these functional links, cognitive functions with posterior probability more than 0.3 had been decoded using neurosynth meta-analysis database. \textbf{(a-d) Histogram and Connectome plots for decoded functions across 12 different emotion conditions:} In every graph for individual function, left diagram is the histogram representing occurrence of cortical regions making connections among each other, right figure shows functional connections on the glass brain (other plots and overlapping functions are shown in supplementary fig S5). \textbf{(e-h)} These subfigures show statistically significant activity in different brain regions across the slices in the z-axis. These statistical maps are obtained from neurosynth meta-analysis database. Using these meta-analytic statistical maps, we were able to find out coordinates which are consistently active across several research studies and labelled these active coordinates with anatomical labels. The heat map of anatomical labelling is shown in \textbf{(i)}. Blue colour shows consistent activation with 99.5\% confidence. Complete labelling against 9 functions(taken on the y-axis in the heat map) and related abbreviations are included in supplementary section 'Active Regions found from Neurosynth Analysis'. Abbreviations: m-medial; l-left; r-right; Mid-Middle; Inf-Inferior; Sup-Superior; TG-Temporal gyrus; PL-Parietal Lobule; AntC-Anterior cingulate; CinG-Cingulate gyrus; FG-Fusiform gyrus; LinG-Lingual gyrus; PosC-Post Central gyrus; Cun-Cuneus; FusG-Fusiform gyrus; Ins-Insula; PreG-Pre Central gyrus.}

\label{fig:Connectome_Plots}
\end{figure}

\begin{figure}[H]
\begin{subfigure}[H]{\textwidth}
\centering
\graphicspath{{pdf/}}
\includegraphics[width=0.7\textwidth, height=0.3\textwidth]{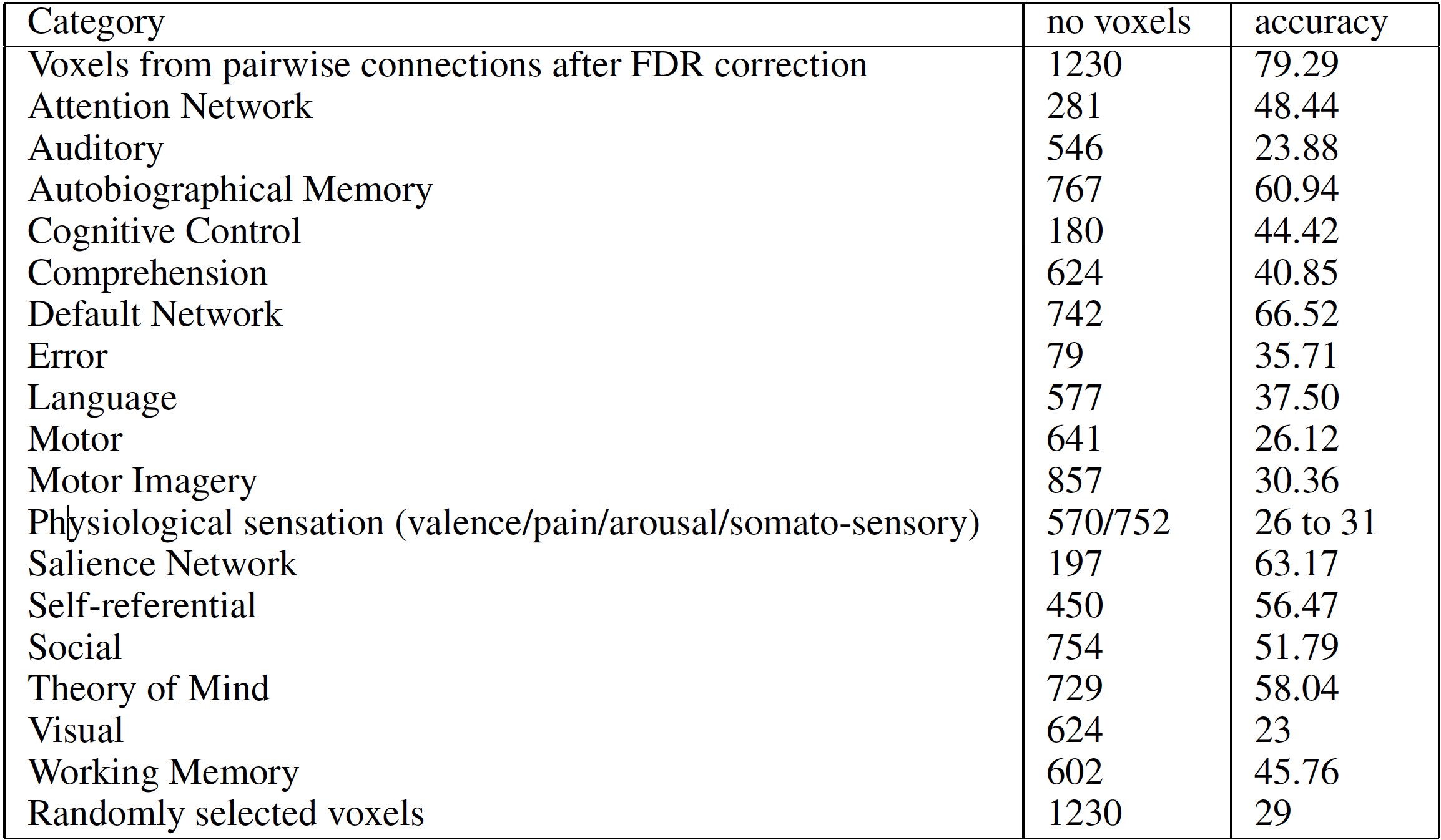}
\caption{{\tiny MVPA analysis using deep learning}} 
\label{fig:mvpatable}
\end{subfigure}

\begin{subfigure}[H]{0.5\textwidth}
\graphicspath{{pdf/Confusion_Matrix/}}
\includegraphics[width=0.9\textwidth, height=0.9\textwidth]{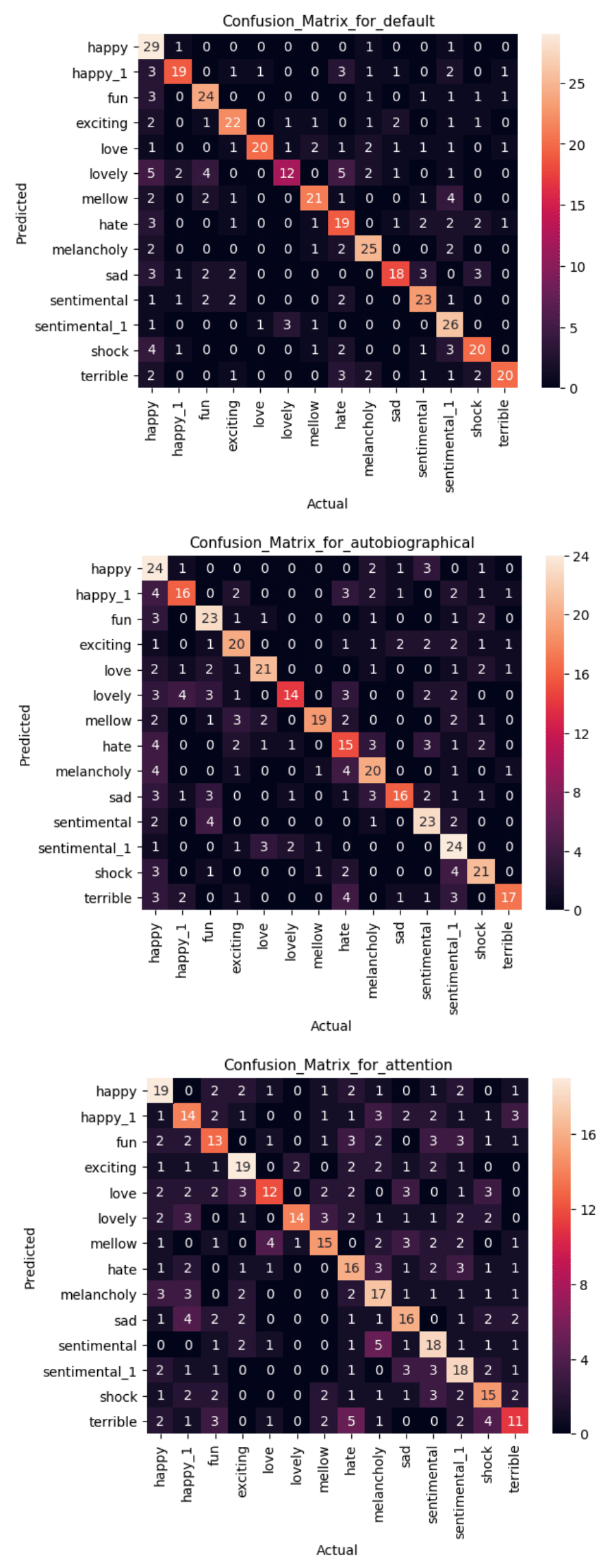}
\caption{{\tiny Confusion Matrix for DMN, AM and AN(top to bottom)}} 
\label{fig:DMN_AM_AN_Conf}
\end{subfigure}
\hfill
\begin{subfigure}[H]{0.5\textwidth}
\centering
\graphicspath{{pdf/ROC/}}
\includegraphics[width=0.9\textwidth, height=0.9\textwidth]{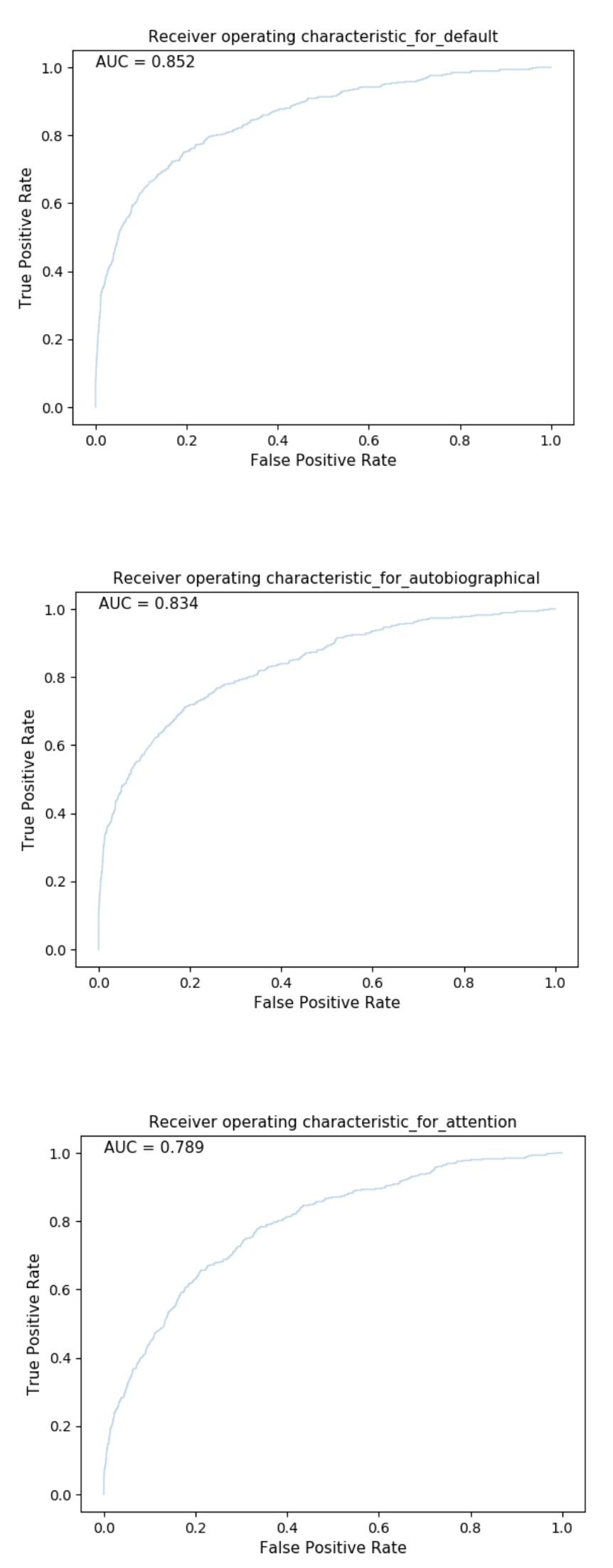}
\caption{{\tiny ROC for DMN, AM and AN(top to bottom)}} 
\label{fig:DMN_AM_AN_ROC}
\end{subfigure}

\begin{subfigure}[H]{0.5\textwidth}
\graphicspath{{pdf/Confusion_Matrix/}}
\includegraphics[width=0.9\textwidth, height=0.9\textwidth]{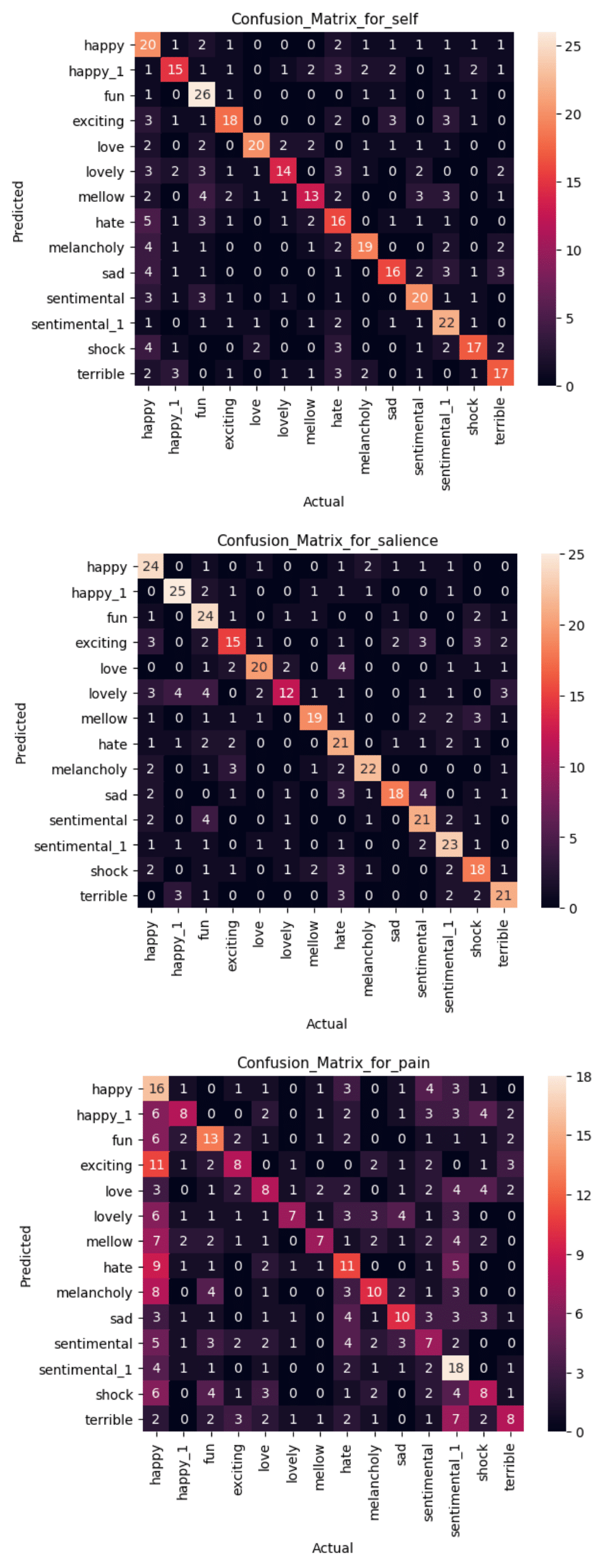}
\caption{{\tiny Confusion Matrix for Self-Referential, Salience Network and Pain(top to bottom)}} 
\label{fig:SelfR_SN_Pain_Conf}
\end{subfigure}
\hfill
\begin{subfigure}[H]{0.5\textwidth}
\centering
\graphicspath{{pdf/ROC/}}
\includegraphics[width=0.9\textwidth, height=0.9\textwidth]{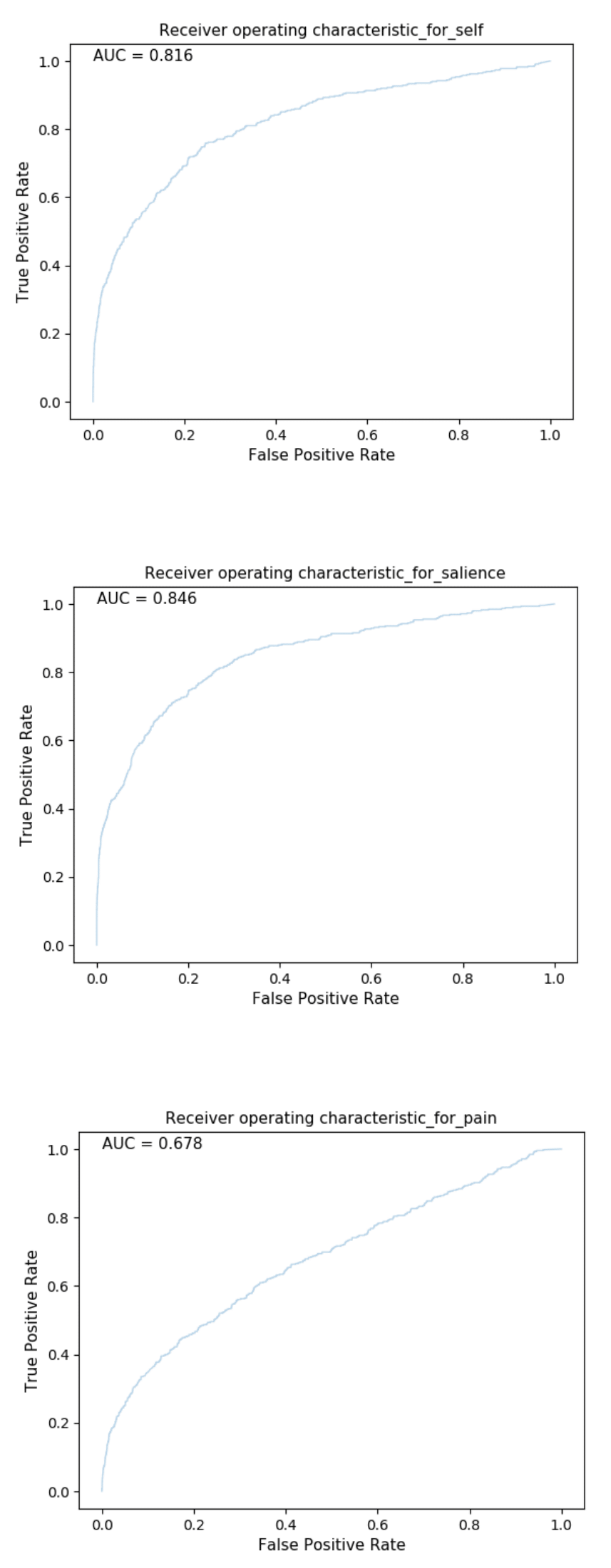}
\caption{{\tiny ROC for Self-Referential, Salience Network and Pain(top to bottom)}} 
\label{fig:SelfR_SN_Pain_ROC}
\end{subfigure}
\end{figure}

\begin{figure}[H]\ContinuedFloat
\begin{subfigure}[H]{0.5\textwidth}
\graphicspath{{pdf/Confusion_Matrix/}}
\includegraphics[width=0.9\textwidth, height=0.9\textwidth]{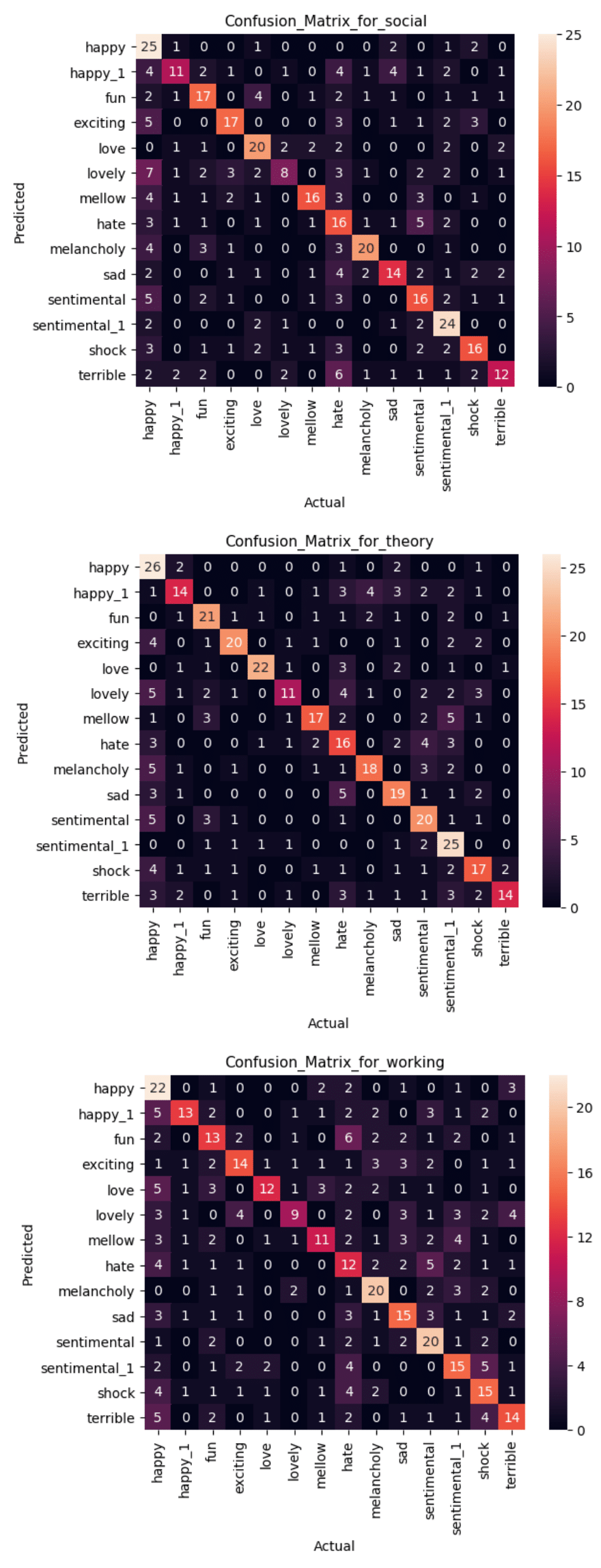}
\caption{{\tiny Confusion Matrix for Social, ToM and Working Memory(top to bottom)}} 
\label{fig:WM_ToM_Social_Conf}
\end{subfigure}
\hfill
\begin{subfigure}[H]{0.5\textwidth}
\centering
\graphicspath{{pdf/ROC/}}
\includegraphics[width=0.9\textwidth, height=0.9\textwidth]{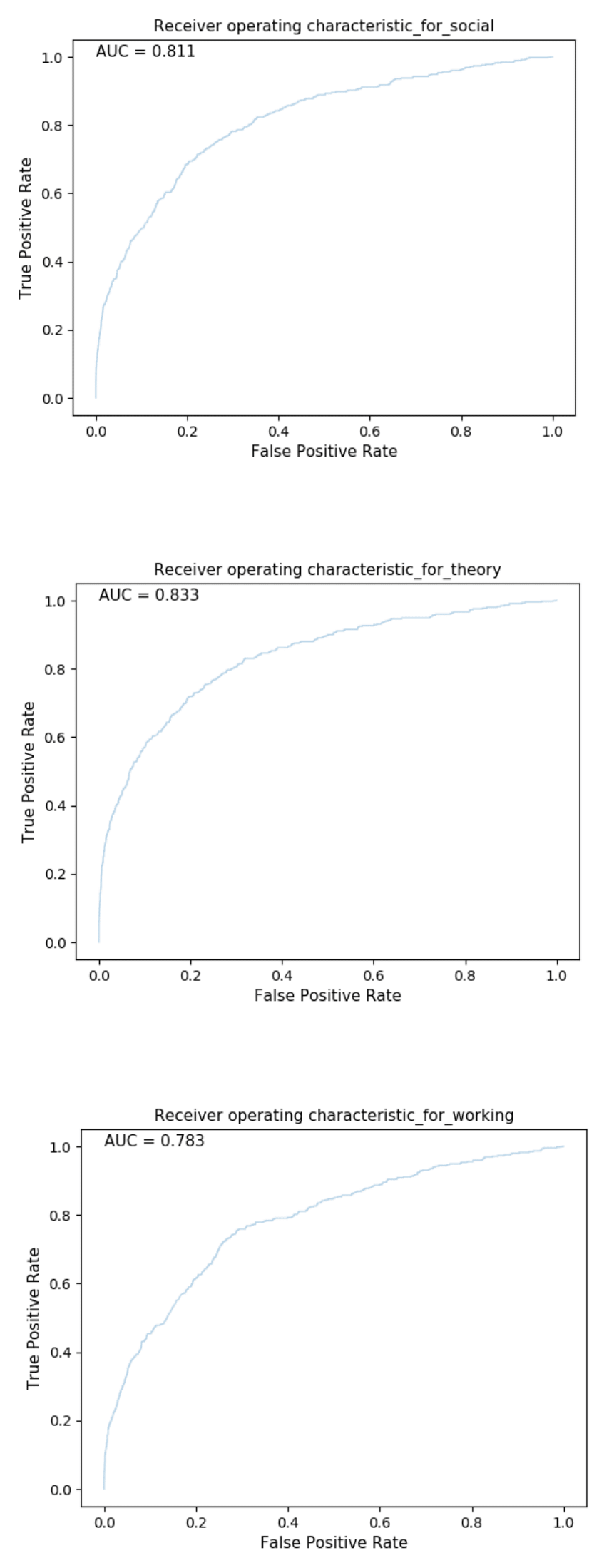}
\caption{{\tiny ROC for Social, ToM and Working Memory(top to bottom)}} 
\label{fig:WM_ToM_Social_ROC}
\end{subfigure}
\hfill
\caption{\textbf{MVPA analysis using deep learning: (a)} Using deep learning(for architecture see fig~\ref{fig:Deep_Model}), we classified emotions on 12 unique categories for 18 functions. Out of 18 different functions, we are getting comparable accuracy for autobiographical memory, default mode network, salience network, theory of mind and self-referential. To cross-check that the set of voxels, which we have calculated and which are associated with the functions mentioned in the table, are not random, any set of random voxels have been picked which gave us only 29\% accuracy. \textbf{(b) Confusion matrix and ROC curve:} Confusion matrices are plotted on the left side, and ROC curves are plotted on the right side. X-axis and Y-axis in the confusion matrix show the target or true classes and predicted classes, respectively. The diagonal elements inside the confusion matrix depict the number of times the predicted class is matched with the true class and non-diagonal elements show the disagreement between true and predicted classes. ROC curve is a probability curve telling about the capability of the model in categorizing the classes. In ROC curve, X and Y axes are false-positive rate(FPR) and true positive rate(TPR), respectively. The area under the curve(AUC) is mentioned inside the plot. More the value of AUC, better the classification model(see fig S11 in supplementary section). Confusion matrix, ROC, loss curve and output of intermediate trained convolution layers for the calculated voxels are shown in fig-S9 (in the supplementary section). For Confusion matrix and ROC curves of other functions, please see fig-S10 (in supplementary section).}
\label{fig:Conf_ROC}
\end{figure}

\begin{figure}[H]
\centering
\graphicspath{{pdf/}}
\includegraphics[width=0.55\textwidth, height=0.55\textwidth]{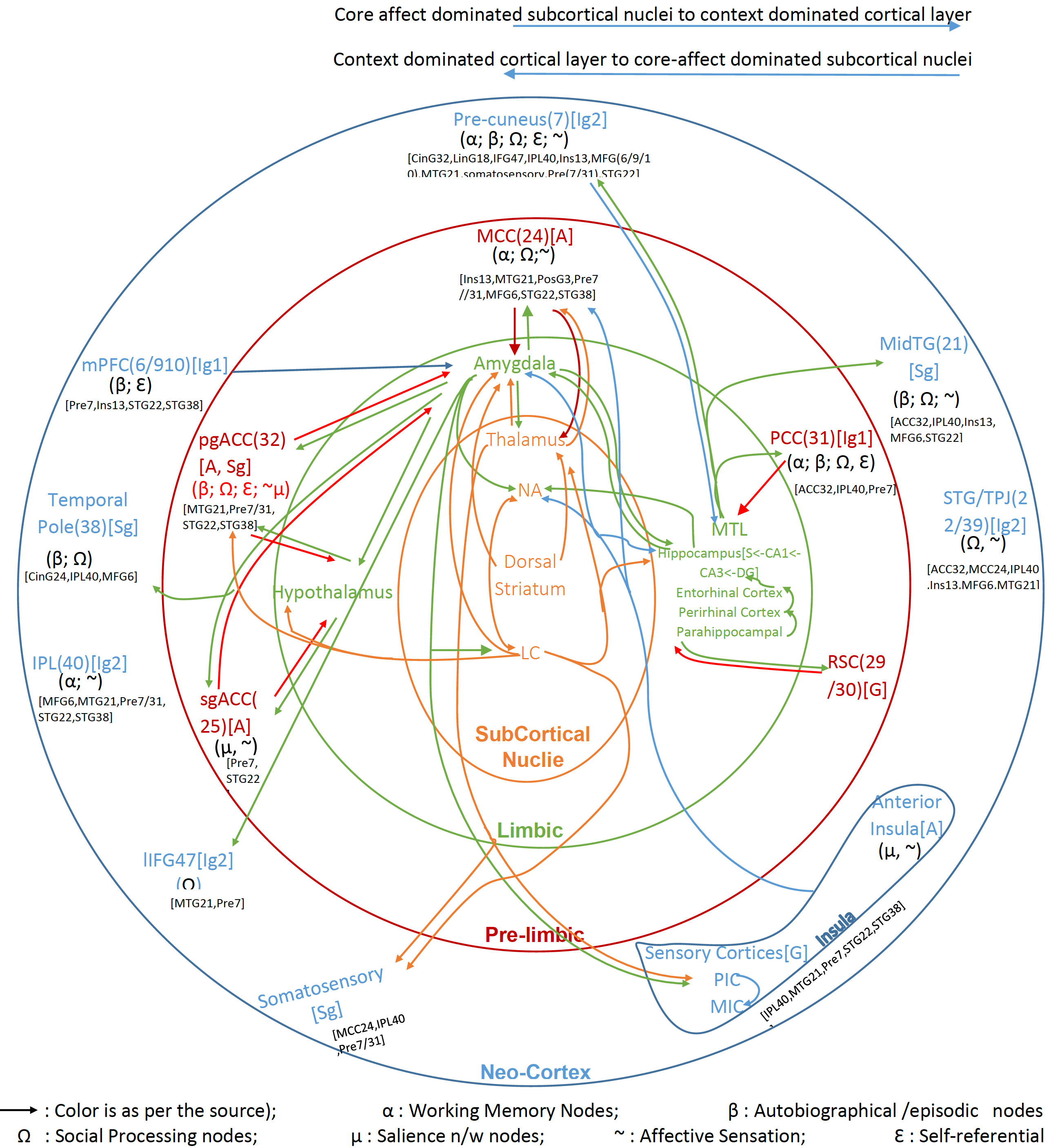}
\caption{\textbf{The cognition-affect integrated model of emotion: } The four-layer structure depicts functional specificity ranging between the core affect specific subcortical nuclei and context processing specific cortical regions. The neocortical regions are connected via long-range connections(\cref{fig:connect_tom,fig:connect_auto,fig:connect_saliency,fig:connect_pain} and supplementary fig S5) and interact with the prelimbic and limbic memory system~\cite{squire1991medial, squire2004medial, eichenbaum2007medial} to process the context. On the other hand, subcortical nuclei react to reward, arousal, pleasure and physiological sensations. These two systems interact with each other in a loop via short-range and long-range projections. This learnt cortico-cortical and cortico-subcortical interaction loop at the local (fig~\ref{fig:Microstructure}) and global scale (fig~\ref{fig:Connectome_Plots}), constructs an integrated state called an emotion. The affect and context can be dissociated at various levels of hierarchical interaction, and information processing (fig~\ref{fig:HierarchicalStructure}) as the context has a structural and temporal hierarchy. Following the hierarchical structure arrangement, there are moments where affect is dominating. In this case, the contextual consideration is minimal and primary, which result in the quick and implicit response (maybe very much related to the survival reflexes). On the other hand, with the varying degree of considerations of this hierarchical structure of the context, varying degree of emotional responses can take place. Since emotion is a learnt concept and associated with the well being, it can be stored for the future reference and recalled as anticipation in order to achieve the allostatic gain. The connections among cortical regions are as per our results and depicted in square brackets. These connections are for the set of cognitive functions including default mode network and salience network, social/ToM, autobiographical memory, physiological sensations, or affect). The subcortical regions were calculated using neurosynth meta-analysis for the affect related functions(valence, arousal, physiological sensation/pain) which we have obtained in our cortical functional network decoding. The interaction between cortex and subcortex, in emotion/affect/sensation condition, is reported in many studies~\cite{koelsch2014brain, janak2015circuits, pessoa2010emotion, cai2018brain, salzman2010emotion, rolls2015limbic, panksepp2011basic, barbas2016prefrontal, fadok2018new, schwarz2015viral, rodenkirch2019locus, zelikowsky2013prefrontal, ye2017direct, tanaka2014cortical, xu2016distinct, barbas1995anatomic}. Abbreviations: NA:Nucleus accummbens ; LC: Locus coeruleus; sg:sub-geneual ; ACC:anterior cingulate cortex ; IPL: Inferior parietal lobule; mPFC: medial pre-frontal cortex; SMA: supplementary motor area; STG: superior temporal gyrus; TPJ: temporo-parietal junction; PIC: posterior insular cortex; MIC: middle insular cortex; IFG: inferior frontal gyrus; MCC: medial cingulate cortex; pg: pregenual; RSC: retrosplenial cortex; PCC: posterior cingulate cortex, MTL: medial temporal lobe. Sg, Ig1, Ig2, and A are representing granularity and described in fig~\ref{fig:Brain_Structure}.} 
\label{fig:model_}
\end{figure}

\end{document}